\documentclass[a4paper,fleqn,usenatbib]{mnras}

\usepackage[T1]{fontenc}
\usepackage{ae,aecompl}
\usepackage{float}
\usepackage{graphicx}   
\usepackage{color}
\usepackage{amssymb}
\usepackage{amsmath}

\title[Photometric and radial-velocity data of RR Lyrae stars in M3]{Photometric and radial-velocity time-series of RR Lyrae stars in M3: analysis of single-mode variables}
\author[J. Jurcsik et al.]{J. Jurcsik$^{1}$\thanks{E-mail: jurcsik@konkoly.hu}, P. Smitola$^{1}$,  G. Hajdu$^{2,3}$, \'A. S\'odor$^{1}$, J. Nuspl$^{1}$, K. Kolenberg$^{4,5}$,  \and G. F\H ur\'esz$^{8}$, L. G. Bal\'azs$^{1}$, C. Pilachowski$^{9}$,  A. Saha$^{10}$, A. Mo\'or$^{1}$,   E. Kun$^{11,12}$, \and A. P\'al$^{1}$, J. Bakos$^{1}$, J. Kelemen$^{1}$, T. Kov\'acs$^{13}$, L. Kriskovics$^{1}$, K. S\'arneczky$^{1}$, \and T. Szalai$^{14}$, A. Szing$^{15}$  K. Vida$^{1}$ \\
\\
$^{1}$Konkoly Observatory of the Hungarian Academy of Sciences, H-1121 Budapest, Konkoly Thege Mikl\'os \'ut 15-17., Hungary\\
$^{2}$Instituto de Astrof\'{i}sica, Pontificia Universidad Cat\'olica de Chile, Av. Vicu\~na Mackenna 4860, 782-0436 Macul, Santiago, Chile\\
$^{3}$Instituto Milenio de Astrof\'isica, Santiago, Chile\\
$^{4}$Institute of Astronomy, KU Leuven, Celestijnenlaan 200D, 3001 Heverlee, Belgium\\
$^{5}$Physics Department, University of Antwerp, Groenenborgerlaan 171, B-2020 Antwerpen, Belgium\\
$^{6}$Institute of Astronomy, KU Leuven, Celestijnenlaan 200D, B-3001 Heverlee, Belgium\\
$^{7}$Harvard-Smithsonian Center for Astrophysics, 60 Garden Street, Cambridge MA 02138, USA\\
$^{8}$MIT Kavli Institute for Astrophysics and Space Research 77 Mass Ave 37-515, Cambridge, MA, 02139\\
$^{9}$Department of Astronomy, Indiana University Bloomington, Swain West 319, 727 E. 3rd Street, Bloomington, IN 47405, US\\
$^{10}$National Optical Astronomy Observatories, Tucson, AZ 85726-6732, USA\\
$^{11}$Department of Theoretical Physics, University of Szeged,  H-6720 Szeged, Tisza Lajos krt 84-86, Hungary\\
$^{12}$Department of Experimental Physics and Astronomical Observatory, University of Szeged, H-6720 Szeged, D\'om t\'er 9, Hungary\\
$^{13}$Institute of Theoretical Physics, Eotvos University, P.O. H-1518, Budapest, Hungary\\
$^{14}$Department of Optics and Quantum Electronics, University of Szeged, H-6720 Szeged, D\'om t\'er 9, Hungary\\
$^{15}$Baja Observatory, University of Szeged, 6500 Baja, KT: 766, Hungary\\}

\begin{document}

\date{Accepted 2017 ..... Received 2017 ..., in original form }

\pagerange{\pageref{firstpage}--\pageref{lastpage}} \pubyear{2016}

\maketitle
\label{firstpage}

\begin{abstract}
We present the first simultaneous photometric and spectroscopic investigation of a large set of RR Lyrae variables in a globular cluster. The  radial-velocity data presented comprise the largest sample of RVs of RR Lyrae stars ever obtained. The target is M3; $BVI_{\mathrm{C}}$ time-series of 111  and $b$ flux data of further 64 RRab stars, and RV data of 79 RR Lyrae stars are published. Blazhko modulation of the light curves of 47\,percent of the RRab stars are detected. The  mean value of the  center-of-mass velocities of RR Lyrae stars is $-146.8$\,km\,s$^{-1}$ with 4.52\,km\,s$^{-1}$ standard deviation, which is in good agreement with the results obtained for the red giants of the cluster. The ${\Phi_{21}}^{\mathrm RV}$  phase difference of the RV curves of RRab stars is found to be uniformly constant both for the M3 and for Galactic field RRab stars; no period or metallicity dependence of the ${\Phi_{21}}^{\mathrm RV}$ is detected.
The Baade-Wesselink distances of 26 non-Blazhko variables with the best phase-coverage radial-velocity curves are determined; the corresponding distance of the cluster, $10480\pm210$ pc,  agrees with the previous literature information. A quadratic formula for the $A_{\mathrm{puls}}-A_V$ relation of RRab stars is given, which is valid for both OoI and OoII variables. We also show that the $(V-I)_0$ of RRab stars measured at light minimum is period dependent, there is at least 0.1 mag difference between the colours at minimum light of the shortest- and longest-period variables.

\end{abstract}


\section{Introduction}\label{sect.int}

Although the history of studying pulsating variables dates back more than a century, these investigations, using the most advanced-technique satellite and ground-based observations (e.g., HST, Kepler, Gaia, LSST, etc.), are still in the focus of modern astrophysics. Exploiting the advantage that globular clusters (GCs) are rich in RR Lyrae (RRL) type pulsating variable stars, and they provide relatively homogeneous samples of variables compared to other targets of massive photometric campaigns (e.g., Galactic field, Galactic bulge, satellite galaxies), GCs are frequent and popular targets of observations. 

In spite of all these efforts, extended radial velocity (RV) surveys of RRL stars in GCs have not been undertaken. Complete RV curves were published for less than 50 Galactic field RRL stars and less than 10 RRLs in GCs previously.  Photometric campaigns long and dense enough to reveal the modulation properties of RRL stars in GCs have not been performed previously, either. Because the origin of the light-curve modulation, i.e. the Blazhko effect, which appears in about 50\,percent of RRLs, is still unknown, a systematic study of Blazhko stars in GCs would provide important information on the Blazhko phenomenon. Therefore, the primary aim of our project was to collect extended multi-colour photometric data, as well as RV observations of the variables in one GC in order to perform a combined analysis of the data. 

The GC, M3 (NGC5272) was selected as the target of the campaign, because it has one of the largest RRL population, and the variables are not too faint to observe spectroscopically. The large number of RRL stars in the cluster motivated many detailed investigations already. Without completeness, \cite{ca05} determined and analysed the physical properties of RRL stars \citep{ca05}, the light curves of the  variables were  modeled accurately using nonlinear pulsation models \citep{marc}, and the comparison of the observed properties of the variables and their distribution with canonical horizontal-branch models and evolutionary tracks led to important new conclusions \citep{mar03,ca4,ccc}. However, in spite of that M3 has  already been the subject of several photometric [e.g., \citet{K98}, \citet{co01}, \citet{cc04} \citet{H05}, and \citet{be06}] and  spectroscopic [e.g., \citet{cm}, \citet{j05}, \citet{sneden} and  \citet{gp}] studies,  the available photometric data provide only uncertain and deficient information on the properties of the Blazhko stars of the cluster, and RV data of the variables has not been published previously.

Utilizing RV and photometric data, the Baade-Wesselink (BW) method \citep[as a review read][]{gautschy} was long one of the most important tools to determine the distance of pulsating variables. Different variants of the method were successfully applied for dozens of Galactic field and some GC RRL stars \citep[without completeness, e.g.,][]{cc90,ccf,c92,f94,lj90,sk89,sk93,st94}.  Although, at the dawn of the Gaia era the determination of distances of clusters through the BW method may well become superseded, the method itself will continue to provide consistent distances of individual pulsating stars at large distances. We point out that for these stars (at distances comparable to M3 and farther away), the relative Gaia parallax error won't allow the derivation of better individual distances than those coming from the BW method. Moreover, the combined, parallel  RV and photometric observations may still provide new information on the pulsation of stars.

In this paper, we present new, extended, multicolour photometric time-series of the fundamental mode RRL stars of M3 and also RV observations of the variables for the first time. The published RV data comprise the largest sample for RRL stars ever obtained. A detailed analysis of the photometric data of the overtone and double-mode variables, based on the same observational campaign, was already published in \cite{overtone}.  The BW analysis of the single-mode variables of M3 is in the focus of the present paper; the study of Blazhko stars will be published separately.

\section{Observations and data reduction}\label{sect.data}

\subsection{Photometry}

Photometric time-series were collected in $BVI_{\mathrm C}$ bands  with the 60/90 Schmidt telescope at Konkoly Observatory in 2012. Both aperture (IRAF/DAOPHOT/PHOT\footnote{IRAF is distributed by the National Optical Astronomy Observatories, which are operated by the Association of Universities for Research in Astronomy, Inc., under cooperative agreement with the National Science Foundation.} ) and Image Subtraction Method \citep[ISIS,][]{isis} photometry of the variables were performed. The light curves were transformed to the standard $BVI_{\mathrm C}$ system  using 23 bright, photometrically stable standard stars, covering a wide range in colour, from the list of \citet{st00}.  Full details of the observations, reduction and the photometry are given in \cite{overtone}, where the analysis of the overtone and the double-mode variables of M3, based on the data from the same photometric campaign, were published.

Standard  $BVI_{\mathrm {C}}$ photometric time-series data of 111 fundamental-mode variables were obtained. Because of crowding, the magnitude calibration of the flux data derived from ISIS failed in the case of  64 RRab stars;  instrumental $B$-band flux time series ($b$) are given for these stars. The $BVI_{\mathrm {C}}$ magnitudes and the $b$ flux time-series of the fundamental-mode variables of M3 are available as Supplementary Material in the electronic edition of this article; Table~\ref{photmag} and Table~\ref{photflux} give samples of their form and content.

\subsection{Spectroscopy}\label{spect}

\begin{table} 
\begin{center} 
\caption{$B$, $V$, $I_{\mathrm{C}}$ CCD observations of M3 variables published in this paper. The full table is available  as Supplementary Material in the online version of the journal. \label{photmag}} 
\begin{tabular}{lccc} 
\hline  
Star$^a$&HJD$-2\,400\,000^b$& Mag & Band \\ 
\hline  
V001& 55937.67595&16.373&B\\ 
V001& 55937.69025&16.321&B\\ 
V001& 55937.70074&16.367&B\\ 
...&  ...    &     ... &     ...\\ 
\hline  
\multicolumn{4}{l}{$^a$Identification according to the catalogue of \cite{samus}.}\\ 
\multicolumn{4}{l}{$^b$The HJDs represent the times of mid-exposure.}
\end{tabular} 
\end{center} 
\end{table} 

\begin{table} 
\begin{center} 
\caption{Instrumental B-band flux time series ($b$) data  of M3 variables with uncertain magnitudes on the reference frame. The complete table is given in the electronic edition.\label{photflux}} 
\begin{tabular}{lcrl} 
\hline  
Star$^a$ & HJD$-2\,400\,000^b$ & Flux$^c$ & \\ 
\hline  
V004a & 55935.64720  &       9768.14&\\ 
V004a & 55937.67595  &   $-$11001.05&\\ 
V004a & 55937.69025  &    $-$8166.03&\\ 
...   &  ...         &       ...    &\\ 
\hline  
\multicolumn{4}{l}{$^a$Identification according to the catalogue of \cite{samus}.}\\ 
\multicolumn{4}{l}{$^b$The HJDs represent the times of mid-exposure.} \\
\multicolumn{4}{l}{$^c$Flux differences derived from ISIS photometry \citep{isis}.}
\end{tabular}
\end{center} 
\end{table} 
 
\begin{table} 
\begin{center} 
\caption{Radial-velocity time-series data of M3 variables. The complete table is given in the electronic edition.\label{rv.dat}} 
\begin{tabular}{lccc} 
\hline  
Star & HJD$-2\,400\,000^{a}$ & RV [km\,s$^{-1}$] & Tel./Instr.$^{b}$  \\ 
\hline  
V009 & 51600.791& $-$185.93 &1\\
V009 & 51600.805& $-$179.69 &1\\
V009 & 51600.818& $-$190.43 &1\\
...&  ...    &     ...   & ... \\  
\hline  
\multicolumn{4}{l}{$^{a}$The HJDs represent the times of mid-exposure.} \\
\multicolumn{4}{l}{$^{b}$ 1:Hydra@WIYN; 2:Hectochelle@MMT.  }
\end{tabular}
\end{center} 
\end{table}

\subsubsection{Hectochelle@MMT measurements}\label{sect.hecto}

Parallel to the photometric measurements, spectroscopic data of M3 RRL stars were secured using the Hectochelle spectrograph attached to the 6.5-m MMT telescope in Arizona, USA \citep{mmt} on 2 nights in March (JD$2455991, 2455992$) and on 3 nights in April (JD$2\,456\,023, 2\,456\,025$ and JD$2\,456\,026$), 2012.

Hectochelle is a multi-object echelle spectrograph capable of observing a single echelle-order of up to 240 point sources simultaneously with $R\approx32\,000$ spectral resolution. The measurements covered the $5150-5300$ \AA\ wavelength range containing the strong Mg triplet lines.  
Besides the target stars, several non-variable stars were also observed for calibration purposes.

The 20-min exposure CCD frames were bias and flat-field corrected, filtered for cosmics, then apertures were extracted and wavelength-calibrated using standard IRAF procedures by the instrument's pipeline. The spectra were then normalised and corrected for the barycentric motion. Finally, we calculated the cross-correlation function (CCF) of the calibrated spectra with a synthetic template F-star spectrum in the range of $[-250\, ,-50]$ km\,s$^{-1}$, and the RVs were  determined by fitting a Gaussian function to the global maximum of the CCF. We checked for systematic uncertainties present in the deduced RV data by calculating the CCFs and RVs in several different ways; applying different templates ($T_{\mathrm {eff}}$: 6500/7000/7250), and selecting different sub-sets of spectral lines.  It was found that these changes affected primarily the derived amplitude of the RV curve; hotter temperature and/or fewer lines gave rise to larger amplitudes. Nevertheless, the differences between the derived RV amplitudes do not exceed 5\,percent of the total amplitude, i.e. it is smaller than 3\,km\,s$^{-1}$ even for the variables with the largest amplitude.

Due to the time consuming-robotic focal-plane fibre-entrance manipulations of Hectochelle, ThAr wavelength reference spectra were obtained only before the first observing night. The RV stability was therefore ensured by relying on the RVs of the reference stars (not standards). However, because the RVs of only 4 reference stars proved to remain stable enough, the RV curves of stable-light-curve RRL stars were also used for RV calibration purposes, and the following co-trending method was applied. The raw pulsation RV curves of all the stable-light-curve RRab stars were pre-whitened by carrying out a preliminary Fourier-fit using pulsation periods determined from photometry. Then, the residuals of each exposure of a subsample of stars of adjacent fibre apertures, selected among the 4 reference stars and the stable-light-curve RRLs, were averaged. Finally, these averaged residuals were subtracted from each target's RV data in the corresponding aperture bin. These corrections were less than $\pm1$\,km\,s$^{-1}$ for the first run of the spectroscopy and were in the $2-4$ \,km\,s$^{-1}$ range for the second run. The 1-$\sigma$ stability of the reference stars and the rms scatter of appropriate order Fourier fits of stable-light-curve RRLs for the corrected RV data is better than 1\,km\,s$^{-1}$.

\subsubsection{Hydra@WIYN measurements}

RVs of RRL stars in M3 were also gathered using the Hydra fibre-feed spectrograph attached to the 3.5-m WIYN telescope at Kitt Peak on 5 nights (between JD 2\,451\,600 and 2\,451\,646) in 2000.  The Hydra configurations included both RRL variables and red-giant-branch stars to provide velocity calibration. The observations and basic spectrum reductions were described in detail by \cite{san}. In brief, the Hydra observations were obtained at a resolving power of $R=3000$ and centered at a wavelength of 5100 \AA.

RVs for all the variables and giants were derived using the IRAF task {\it fxcor}.  Velocities were determined by cross-correlation relative to the twilight-motr sky spectrum observed with the same instrument. The spectral region for cross-correlation was restricted to $5000-5400$ \AA\ to avoid H$\beta$. For the giants, the Tonry \& Davis (1979) ``R'' value was  typically above 15, while the RRL variables,  with weaker lines and lower signal-to-noise ratio spectra, had values in the range 5$<$R$<$20.
For the (non-variable) giants, the observed velocity uncertainties are typically 1\,km\,s$^{-1}$ based on the reproducibility of the measurements from observation to observation.   The  uncertainties of the RVs of the variables are estimated from the scatter about the Fourier fit of the RV curves of stable light-curve RRLs; the rms of these RV curves are $5-7$\,km\,s$^{-1}$.
 
In each multi-object Hydra frame, the velocities of the giants were averaged to produce a mean giant velocity to establish the absolute calibration for RV. An offset was added to the measured RRL velocities in that frame to bring the velocities to the same systemic velocity defined by the giant stars. With a relatively small number of giants (only 14) to establish the cluster systemic velocity, our results are subject to sampling errors. These errors are estimated by considering randomly selected samples of 14 giants from the \cite{sod} sample of 88 giants. The standard deviation of the mean velocity of the sub-samples compared to the mean of the full sample of 88 giants is 0.8\,km\,s$^{-1}$.  We take this as an estimate of the uncertainty  of the RV zero-point due to sampling error introduced by our small sample of giants.

\begin{figure}
\centering
\includegraphics[width=9cm]{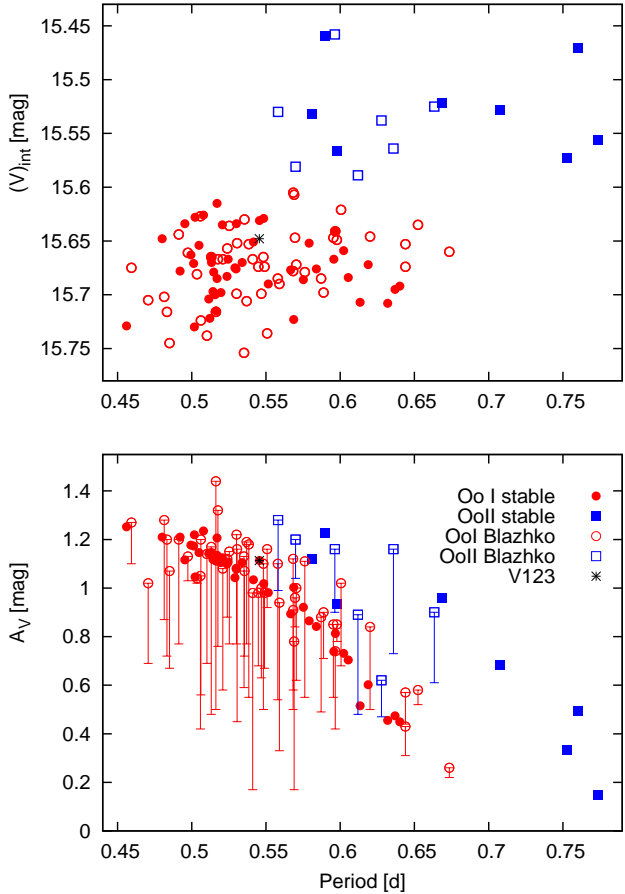}
\caption{Intensity-averaged mean $V$ magnitude and peak to peak $V$ amplitude vs. pulsation period of RRab stars in M3 are plotted. Blazhko stars are shown at the largest amplitude, and their amplitude ranges are indicated by vertical bars. Beside the 2012 data, all the published CCD observations of Blazhko stars are checked in determining minimum/maximum  values of their amplitudes.    \label{plotab}}
\end{figure}

\begin{figure*}
\centering
\includegraphics[width=17.8cm]{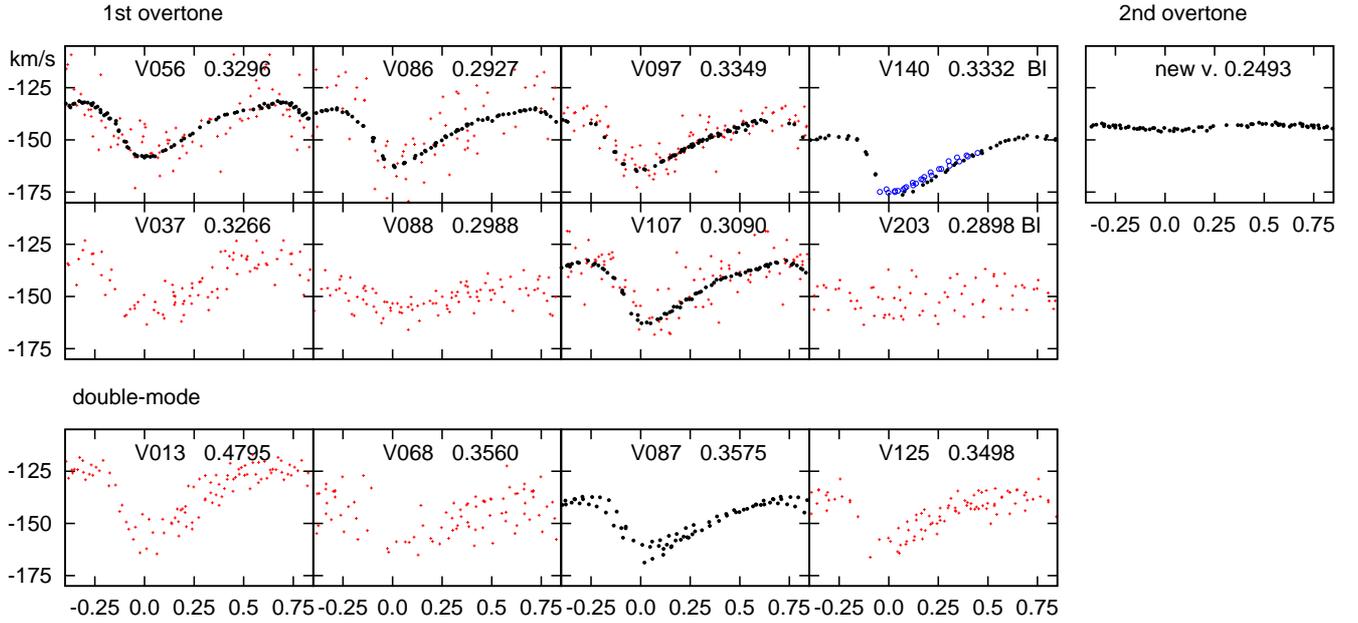}
\caption{Radial-velocity curves of overtone and double-mode RR Lyrae stars in M3. The  Hectoechelle@MMT and the  Hydra@WIYN RV data are shown by black dots ans red crosses, respectively. The radial velocity data of the first (black dots) and second  run (blue circles) of the Hectoechelle@MMT observations for the Blazhko star, V140, are shown by different symbols. The IDs  and the pulsation period (days) of the variables are also given in the panels.\label{rvc}}
\end{figure*}

\begin{figure*}
\centering
\includegraphics[width=18cm]{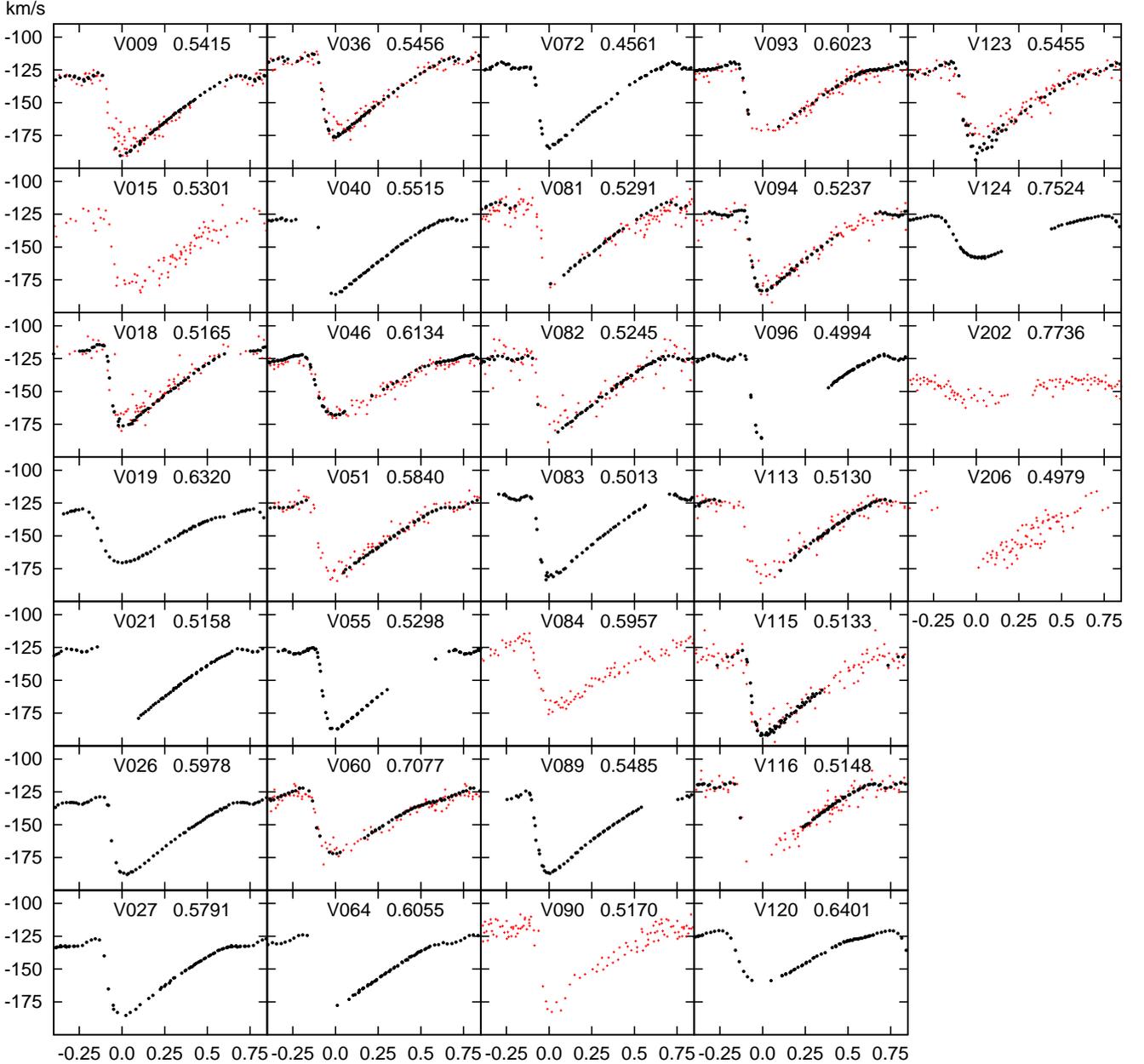}
\caption{Radial-velocity curves of stable-light-curve RRab stars in M3. Symbols and labels are  as given in the caption of  Fig.~\ref{rvc}.\label{rvab}}
\end{figure*}

\begin{figure*}
\centering
\includegraphics[width=18cm]{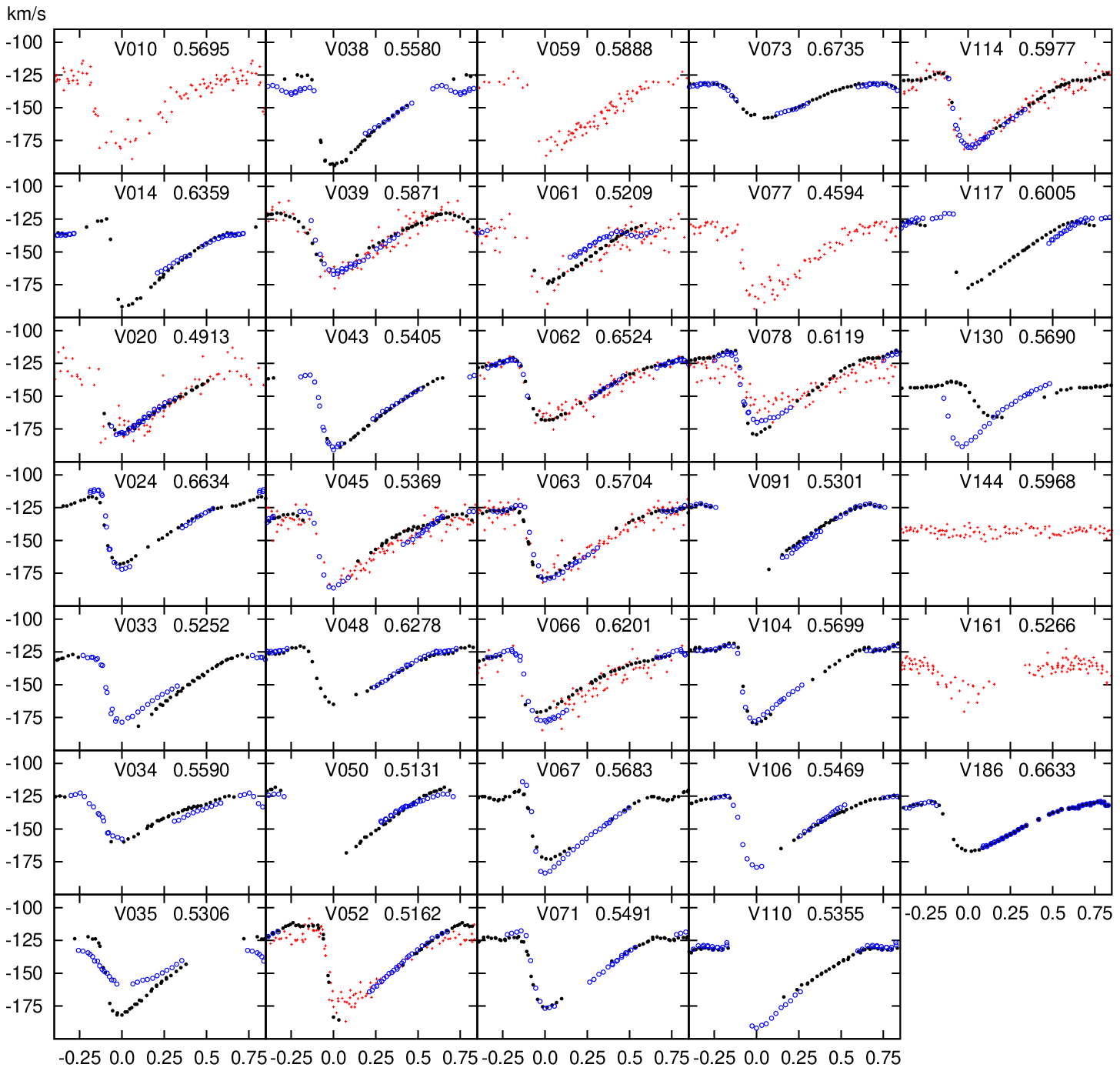}
\caption{Radial-velocity curves of Blazhko RRab stars in M3. Symbols and labels are as given in the caption of  Fig.~\ref{rvc}.\label{rvbl}}
\end{figure*}

\subsubsection{The RV data}

There are 16 non-Blazhko, single-mode, RRL stars with good pulsation-phase coverage in both RV data sets. The differences between the mean RVs ($\gamma$-velocities) of these stars measured by the Hydra@WIYN and the Hectochelle@MMT instruments are in the [$-1.5, 2.8$]\,km\,s$^{-1}$ range, with 0.6 mean value and 1.5\,rms scatter, indicating that there is no systematic difference between the two data sets exceeding the $\sim1$\,km\,s$^{-1}$ uncertainty. 

Altogether 60 and $60-80$ RV measurements  of  62 and 44 RRL stars were obtained by the Hectochelle@MMT and Hydra@WIYN instruments, respectively. Both data sets are available as Supplementary Material in the electronic edition of this article in a merged form.  Table~\ref{rv.dat}  supplies a sample for  this data file showing its structure and the meaning of the columns.

\section{Basic data of the variables}

The light and RV curves were analysed using  the program packages MUFRAN \citep{mufran} and LCfit \citep{nl}.

\subsection{Photometric results}

\begin{table*}
 \begin{minipage}{\textwidth}
\begin{center} 
\caption{Summary of the photometric properties of  RRab stars and the new overtone variable in M3.\label{rrab.dat}} 
\begin{tabular}{l@{\hspace{2mm}}l@{\hspace{3mm}}c@{\hspace{1mm}}c@{\hspace{2mm}}l@{\hspace{1mm}}l@{\hspace{3mm}}c@{\hspace{1mm}}c@{\hspace{2mm}}l@{\hspace{1mm}}l@{\hspace{3mm}}c@{\hspace{1mm}}c@{\hspace{2mm}}l@{\hspace{1mm}}l@{\hspace{8mm}}l}
\hline
Star&Period &$<B>^{a}$ &($B$)$^a$&\multicolumn{2}{c}{$A_B^b$} &$<V>$&($V$)&\multicolumn{2}{l}{\hskip 6mm $A_V^{b,c}$} &$<I_{\mathrm C}>$& ($I_{\mathrm C}$)&\multicolumn{2}{c}{${A_{I_{\mathrm C}}}^b$} & Comment\\
 &\multicolumn{1}{c}{d} &mag&mag&mag&mag&mag&mag&mag&mag&mag&mag&mag&mag&\multicolumn{1}{l}{}  \\ 
\hline
V001 OoI   & 0.52059  & 16.022 & 15.916 & 1.452 &      & 15.694 & 15.635 & 1.125 &      & 15.229 & 15.208 & 0.750 &      & \\
V003 OoII  & 0.55820  & 15.908 & 15.803 & 1.62  & 1.36 & 15.588 & 15.530 & 1.28  & 0.99 & 15.129 & 15.106 & 0.85  & 0.68 & Bl \\
V004a       & 0.58504  & & & & & & & & & \\
V004b       & 0.59305  &&&&&&&&&&&&& Bl \\    
V005 OoI   & 0.50579  & 15.974 & 15.932 & 1.38 & 0.60 & 15.650 & 15.627 & 1.05 & 0.42 & 15.202 & 15.193 & 0.65 & 0.28 &  Bl \\ 
V006 OoI   & 0.51434  & 16.088 & 15.990 & 1.421 &  & 15.751 & 15.697 & 1.118 & &  15.284 & 15.263 & 0.744 & & \\
V007 OoI   & 0.49742  & 16.052 & 15.943 & 1.50 & 1.38 & 15.718 & 15.661 & 1.13 & 1.03 & 15.260 & 15.238 & 0.77 & 0.65 & Bl \\ 
V008        & 0.63671  &&&&&&&&&&&&& Bl \\  
V009 OoI   & 0.54155  & 16.045 & 15.959 & 1.334 &  & 15.697 & 15.651 & 1.034 & &  15.210 & 15.192 & 0.690 & & \\
V010 OoI   & 0.56955  & 16.042 & 15.981 & 1.23 & 1.06 & 15.680 & 15.647 & 0.96 & 0.84 & 15.169 & 15.157 & 0.64 & 0.53 & Bl \\ 
V011 OoI   & 0.50789  & 15.990 & 15.872 & 1.559 &  & 15.694 & 15.626 & 1.235 & &  15.236 & 15.210 & 0.821 & & \\
V014 OoII  & 0.63590  & 15.960 & 15.892 & 1.46 & 0.97 & 15.602 & 15.564 & 1.16 & 0.73 & 15.107 & 15.092 & 0.76 & 0.50 & Bl \\ 
V015 OoI   & 0.53009  & 16.021 & 15.930 & 1.382 &  & 15.684 & 15.634 & 1.078 & &  15.202 & 15.183 & 0.730 & & \\
V016 OoI   & 0.51149  & 16.100 & 15.998 & 1.462 &  & 15.760 & 15.704 & 1.142 & &  15.288 & 15.266 & 0.761 & & \\
V017 OoI   & 0.57613  & 16.058 & 16.016 & 1.15& 0.72 & 15.702 & 15.679 & 0.86 (1.11) & 0.55 & 15.194 & 15.185 & 0.57 & 0.30 & Bl \\ 
V018 OoI   & 0.51645  & 16.106 & 16.011 & 1.421 &  & 15.767 & 15.715 & 1.111 & &  15.295 & 15.274 & 0.749 & & \\
V019 OoI   & 0.63198  & 16.122 & 16.103 & 0.598 &  & 15.718 & 15.708 & 0.455 & &  15.168 & 15.164 & 0.302 & & \\
V020 OoI   & 0.49126  & 15.998 & 15.916 & 1.53 & 1.00 & 15.691 & 15.644 & 1.20 & 0.77 & 15.248 & 15.230 & 0.76 & 0.45 & Bl \\ 
V021 OoI   & 0.51577  & 16.103 & 16.007 & 1.437 &  & 15.753 & 15.700 & 1.115 & &  15.278 & 15.258 & 0.741 & & \\
V022 OoI   & 0.48143  & 16.042 & 15.948 & 1.60 & 1.15 & 15.757 & 15.702 & 1.28 & 0.87 & 15.337 & 15.316 & 0.86 & 0.55 & Bl \\ 
V023 OoI   & 0.59538  & 16.012 & 15.972 & 1.10 & 0.70 & 15.669 & 15.647 & 0.85 & 0.55 & 15.137 & 15.129 & 0.56 & 0.36 & Bl \\ 
V024 OoII  & 0.66343  & 15.953 & 15.908 & 1.18 & 0.78 & 15.549 & 15.525 & 0.90 & 0.61 &        &        &      &      & Bl \\ 
V025 OoI   & 0.48007  & 16.054 & 15.932 & 1.550 &  & 15.714 & 15.648 & 1.210 & &  15.271 & 15.244 & 0.807 & & \\
V026 OoII  & 0.59774  & 15.959 & 15.892 & 1.208 &  & 15.602 & 15.566 & 0.937 & &  15.093 & 15.079 & 0.621 & & \\
V027 OoI   & 0.57906  & 16.048 & 15.994 & 1.116 &  & 15.682 & 15.652 & 0.865 & &  15.200 & 15.188 & 0.576 & & \\
V028 OoI   & 0.47066  & 15.978 & 15.921 & 1.28 & 1.12 & 15.741 & 15.705 & 1.02 & 0.88 (0.69) & 15.312 & 15.298 & 0.66 & 0.55 & Bl \\ 
V030        & 0.51209  & & & & & & & & &   \\
V031 OoII  & 0.58072  & 15.910 & 15.815 & 1.445 &  & 15.584 & 15.532 & 1.118 & &  15.145 & 15.123 & 0.766 & & \\
V032 OoI   & 0.49535  & 16.030 & 15.924 & 1.455 &  & 15.691 & 15.634 & 1.116 & &  15.276 & 15.253 & 0.754 & & \\
V033 OoI   & 0.52524  & 15.996 & 15.926 & 1.50 & 1.06 & 15.674 & 15.636 & 1.15 & 0.77 & 15.190 & 15.176 & 0.77 & 0.46 & Bl \\ 
V034 OoI   & 0.55901  & 16.061 & 16.014 & 1.25 & 0.50 & 15.714 & 15.690 & 0.94 & 0.33 & 15.219 & 15.209 & 0.65 & 0.19 & Bl \\ 
V035 OoI   & 0.53058  & 16.010 & 15.958 & 1.53 & 0.58 & 15.681 & 15.652 & 1.16 & 0.45 & 15.204 & 15.193 & 0.74 & 0.26 & Bl  \\ 
V036 OoI   & 0.54560  & 16.048 & 15.950 & 1.457 &  & 15.684 & 15.631 & 1.114 & &  15.212 & 15.190 & 0.755 & & \\
V038 OoI   & 0.55801  & 16.062 & 16.008 & 1.50 & 0.71 & 15.715 & 15.685 & 1.10 & 0.54 & 15.215 & 15.204 & 0.70 & 0.37 & Bl  \\   
V039 OoI   & 0.58710  & 16.066 & 16.030 & 1.13 & 0.70 & 15.704 & 15.685 & 0.88 & 0.49 & 15.190 & 15.182 & 0.57 & 0.30 & Bl  \\ 
V040 OoI   & 0.55154  & 16.088 & 16.015 & 1.274 &  & 15.730 & 15.690 & 0.981 & &  15.234 & 15.219 & 0.658 & & \\
V041 OoI   & 0.48504  & 16.078 & 15.995 & 1.33 & 1.17 & 15.793 & 15.745 & 1.07 & 0.90 (0.67) &  15.356 & 15.338 & 0.70 & 0.57 & Bl \\ 
V042 OoII  & 0.59007  & 15.869 & 15.749 & 1.609 &  & 15.524 & 15.459 & 1.226 & &  15.063 & 15.037 & 0.816 & & \\
V043        & 0.54053  &&&&&&&&&&&&& Bl \\ 
V045 OoI   & 0.53691  & 16.092 & 16.021 & 1.54 & 1.00 & 15.745 & 15.706 & 1.19 & 0.77 & 15.260 & 15.245 & 0.80 & 0.46  & Bl \\ 
V046 OoI   & 0.61338  & 16.109 & 16.086 & 0.692 &  & 15.719 & 15.707 & 0.515 & &  15.173 & 15.168 & 0.345 & & \\
V047 OoI   & 0.54107  & 16.000 & 15.967 & 1.25 & 0.20 & 15.684 & 15.667 & 0.98 & 0.17 & 15.207 & 15.201 & 0.65 & 0.13 &   Bl \\ 
V048 OoI   & 0.62783  & 15.942 & 15.918 & 0.80 & 0.60 & 15.551 & 15.538 & 0.62 & 0.47 & 14.994 & 14.989 & 0.40 & 0.33 & Bl  \\ 
V049 OoI   & 0.54821  & 16.058 & 15.977 & 1.48 & 1.17 & 15.710 & 15.665 & 1.10 & 0.90 (0.50) & 15.233 & 15.215 & 0.75 & 0.58 & Bl \\  
V050 OoI   & 0.51309  & 16.041 & 15.954 & 1.53 & 1.04 & 15.713 & 15.665 & 1.17 & 0.78 (0.48) & 15.233 & 15.215 & 0.78 & 0.53 & Bl \\ 
V051 OoI   & 0.58397  & 16.090 & 16.036 & 1.105 &  & 15.705 & 15.676 & 0.842 & &  15.192 & 15.180 & 0.576 & & \\
V052 OoI   & 0.51624  & 16.096 & 16.015 & 1.88 & 0.95 & 15.761 & 15.716 & 1.44 & 0.69 (0.50) & 15.285 & 15.267 & 0.96 & 0.46 & Bl \\ 
V053 OoI   & 0.50488  & 16.074 & 15.961 & 1.530 &  & 15.713 & 15.654 & 1.146 & &  15.263 & 15.239 & 0.798 & & \\
V054 OoI   & 0.50611  & 16.072 & 16.032 & 1.57 & 0.78 & 15.748 & 15.724 & 1.20 & 0.56 & 15.310 & 15.300 & 0.82 & 0.27  &  Bl \\ 
V055 OoI   & 0.52983  & 16.064 & 15.972 & 1.380 &  & 15.727 & 15.676 & 1.081 & &  15.232 & 15.213 & 0.718 & & \\
V057 OoI   & 0.51218  & 16.101 & 16.001 & 1.435 &  & 15.779 & 15.722 & 1.145 & &  15.318 & 15.296 & 0.758 & & \\
V058 OoI   & 0.51705  & 16.026 & 15.902 & 1.594 &  & 15.681 & 15.615 & 1.206 & &  15.231 & 15.205 & 0.803 & & \\
V059 OoI   & 0.58882  & 16.095 & 16.045 & 1.16 & 0.90 & 15.725 & 15.698 & 0.90 & 0.71 & 15.204 & 15.194 & 0.59 & 0.40 & Bl  \\ 
V060 OoII  & 0.70773  & 15.947 & 15.909 & 0.922 &  & 15.547 & 15.528 & 0.683 & &  15.023 & 15.015 & 0.448 & & \\
V061$^*$ OoI& 0.52091 & 16.021 & 15.958 & 1.44 & 0.77 & 15.702 & 15.667 & 1.08 & 0.58 & 15.231 & 15.217 & 0.67 & 0.36 & Bl/dm  \\  
V062 OoI   & 0.65240  & 16.046 & 16.022 & 0.77 & 0.69 & 15.648 & 15.635 & 0.58 & 0.52 & 15.097 & 15.092 & 0.37 & 0.33 & Bl \\ 
V063 OoI   & 0.57040  & 16.054 & 16.002 & 1.30 & 0.82 & 15.701 & 15.672 & 1.00 & 0.62 & 15.188 & 15.178 & 0.66 & 0.38 & Bl  \\ 
V064 OoI   & 0.60546  & 16.084 & 16.045 & 0.942 &  & 15.705 & 15.684 & 0.704 & &  15.165 & 15.157 & 0.472 & & \\
V065 OoII  & 0.66835  & 15.922 & 15.856 & 1.229 &  & 15.559 & 15.522 & 0.959 & &  15.041 & 15.026 & 0.629 & & \\
V066 OoI   & 0.62015  & 16.048 & 16.016 & 1.06 & 0.63 & 15.662 & 15.646 & 0.84 & 0.50 & 15.121 & 15.115 & 0.52 & 0.30 & Bl  \\ 
V067 OoI   & 0.56832  & 16.062 & 16.012 & 1.47 & 0.66 & 15.705 & 15.678 & 1.12 & 0.50 & 15.195 & 15.184 & 0.75 & 0.32 & Bl  \\ 
V069 OoI   & 0.56662  & 16.085 & 16.025 & 1.187 &  & 15.710 & 15.677 & 0.894 & &  15.196 & 15.184 & 0.600 & & \\
\end{tabular}
\end{center}
\end{minipage}
\end{table*}

\begin{table*}
\begin{minipage}{\textwidth}
\begin{center}
\contcaption{}
\begin{tabular}{l@{\hspace{2mm}}l@{\hspace{3mm}}c@{\hspace{1mm}}c@{\hspace{2mm}}l@{\hspace{1mm}}l@{\hspace{3mm}}c@{\hspace{1mm}}c@{\hspace{2mm}}l@{\hspace{1mm}}l@{\hspace{3mm}}c@{\hspace{1mm}}c@{\hspace{2mm}}l@{\hspace{1mm}}l@{\hspace{8mm}}l}
\hline
Star  &Period &$<B>^{a}$ &($B$)$^a$&\multicolumn{2}{c}{$A_B^b$} &$<V>$&($V$)&\multicolumn{2}{c}{\hskip 6mm $A_V^{b,c}$} &$<I_{\mathrm C}>$& ($I_{\mathrm C}$)      &\multicolumn{2}{c}{${A_{I_{\mathrm C}}}^b$} & Comment\\
 & \multicolumn{1}{c}{d}&mag&mag&mag&mag&mag&mag&mag&mag&mag&mag&mag&mag&\multicolumn{1}{l}{}  \\ 
\hline
V071 OoI   & 0.54905  & 16.054 & 15.999 & 1.25 & 0.80 & 15.706 & 15.674 & 0.98 & 0.67 & 15.206 & 15.195 & 0.63 & 0.43 & Bl  \\ 
V072 OoI   & 0.45608  & 16.115 & 15.982 & 1.611 &  & 15.803 & 15.729 & 1.252 & &  15.378 & 15.349 & 0.841 & & \\
V073 OoI   & 0.67349  & 16.065 & 16.059 & 0.36 & 0.28 & 15.663 & 15.660 & 0.26 & 0.22 & 15.104 & 15.103 & 0.16 & 0.14 & Bl \\ 
V074 OoI   & 0.49215  & 16.097 & 15.976 & 1.565 &  & 15.744 & 15.678 & 1.210 & &  15.315 & 15.288 & 0.849 & & \\
V076 OoI   & 0.50176  & 16.105 & 15.986 & 1.568 &  & 15.797 & 15.730 & 1.219 & &  15.335 & 15.308 & 0.823 & & \\
V077 OoI   & 0.45935  & 16.057 & 15.931 & 1.70 & 1.50 & 15.742 & 15.675 & 1.27 & 1.10 & 15.365 & 15.337 & 0.90 & 0.76 & Bl  \\ 
V078 OoII? & 0.61192  & 15.985 & 15.947 & 1.21 & 0.68 & 15.609 & 15.589 & 0.89 & 0.48 & 15.059 & 15.051 & 0.61 & 0.33 & Bl \\ 
V079 OoI   & 0.48329  & 16.065 & 15.998 & 1.58 & 0.93 & 15.755 & 15.716 & 1.20 & 0.72 & 15.306 & 15.291 & 0.77 & 0.49 & Bl  \\ 
V080 OoI   & 0.5384  & 16.048 & 15.966 & 1.50 & 1.20 & 15.700 & 15.653 & 1.18 & 0.92 (0.55) & 15.222 & 15.203 & 0.77 & 0.59 & Bl  \\ 
V081 OoI   & 0.52912  & 16.057 & 15.971 & 1.357 &  & 15.722 & 15.675 & 1.043 & &  15.235 & 15.217 & 0.695 & & \\
V082 OoI   & 0.52454  & 16.053 & 15.957 & 1.399 &  & 15.721 & 15.667 & 1.107 & &  15.240 & 15.219 & 0.743 & & \\
V083 OoI   & 0.50127  & 16.061 & 15.951 & 1.494 &  & 15.733 & 15.671 & 1.173 & &  15.276 & 15.252 & 0.793 & & \\
V084 OoI   & 0.59572  & 16.070 & 16.028 & 0.955 &  & 15.689 & 15.667 & 0.739 & &  15.153 & 15.144 & 0.486 & & \\
V089 OoI   & 0.54848  & 16.029 & 15.951 & 1.313 &  & 15.671 & 15.629 & 1.018 & &  15.185 & 15.169 & 0.661 & & \\
V090 OoI   & 0.51703  & 16.074 & 15.974 & 1.438 &  & 15.741 & 15.685 & 1.135 & &  15.264 & 15.242 & 0.756 & & \\
V091 OoI   & 0.53013  & 16.072 & 15.988 & 1.38 & 1.03 & 15.752 & 15.699 & 1.22 & 0.77 & 15.213 & 15.197 & 0.75 & 0.46 & Bl \\ 
V092 OoI   & 0.50355  & 16.055 & 15.966 & 1.33 & 1.32 & 15.731 & 15.681 & 1.04 & 1.02 & 15.265 & 15.247 & 0.66 & 0.64 & Bl  \\ 
V093 OoI   & 0.60230  & 16.053 & 16.012 & 0.954 &  & 15.681 & 15.659 & 0.731 & &  15.143 & 15.134 & 0.471 & & \\
V094 OoI   & 0.52370  & 16.075 & 15.981 & 1.398 &  & 15.735 & 15.683 & 1.099 & &  15.257 & 15.236 & 0.739 & & \\
V096 OoI   & 0.49941  & 16.038 & 15.925 & 1.509 &  & 15.726 & 15.663 & 1.177 & &  15.248 & 15.224 & 0.769 & & \\
V100 OoI   & 0.61882  & 16.082 & 16.052 & 0.809 &  & 15.687 & 15.672 & 0.602 & &  15.173 & 15.166 & 0.415 & & \\
V101 OoI   & 0.64389  & 16.099 & 16.076 & 0.79 & 0.58 & 15.686 & 15.674 & 0.57 & 0.44 & 15.136 & 15.131 & 0.38 & 0.28 & Bl  \\ 
V104 OoII  & 0.56993  & 15.980 & 15.878 & 1.55 & 1.38 & 15.635 & 15.581 & 1.20 & 1.04 & 15.162 & 15.141 & 0.80 & 0.69 & Bl  \\ 
V106 OoI   & 0.54689  & 16.061 & 16.009 & 1.25 & 0.85 & 15.729 & 15.699 & 1.00 & 0.63 & 15.260 & 15.248 & 0.68 & 0.37 & Bl  \\ 
V108 OoI   & 0.51961  & 16.086 & 15.991 & 1.408 &  & 15.752 & 15.698 & 1.105 & &  15.275 & 15.255 & 0.750 & & \\
V109 OoI   & 0.53391  & 16.078 & 15.985 & 1.412 &  & 15.721 & 15.670 & 1.105 & &  15.223 & 15.204 & 0.730 & & \\
V110 OoI   & 0.53547  & 16.010 & 15.942 & 1.43 & 0.94 & 15.665 & 15.630 & 1.07 & 0.72 & 15.206 & 15.192 & 0.73 & 0.48 & Bl  \\ 
V111 OoI   & 0.51019  & 16.079 & 16.014 & 1.51 & 0.96 & 15.772 & 15.738 & 1.14 & 0.69 & 15.270 & 15.256 & 0.75 & 0.44 & Bl \\ 
V113 OoI   & 0.51301  & 16.055 & 15.952 & 1.460 &  & 15.723 & 15.665 & 1.143 & &  15.260 & 15.237 & 0.772 & & \\
V114 OoI   & 0.59773  & 16.046 & 15.997 & 1.09 & 1.00 & 15.676 & 15.649 & 0.85 & 0.78 & 15.172 & 15.162 & 0.53 & 0.49 & Bl \\ 
V115 OoI   & 0.51335  & 16.055 & 15.953 & 1.448 &  & 15.728 & 15.670 & 1.146 & &  15.247 & 15.225 & 0.764 & & \\
V116 OoI   & 0.51481  & 16.067 & 15.966 & 1.436 &  & 15.734 & 15.679 & 1.120 & &  15.264 & 15.242 & 0.747 & & \\
V117 OoI   & 0.60054  & 16.005 & 15.950 & 1.32 & 0.90 & 15.652 & 15.621 & 1.02 & 0.68 & 15.137 & 15.126 & 0.67 & 0.42 & Bl \\
V118$^{**}$ OoI&0.49941 & & & & & & & & & \\
V119 OoI   & 0.51758  & 16.051 & 15.969 & 1.60 & 1.05 & 15.711 & 15.667 & 1.32 & 0.76 & 15.247 & 15.230 & 0.87 & 0.46 & Bl \\ 
V120 OoI   & 0.64014  & 16.088 & 16.072 & 0.572 &  & 15.701 & 15.692 & 0.449 & &  15.137 & 15.134 & 0.302 & & \\
V121 OoI   & 0.53520  & 16.116 & 16.063 & 1.46 & 0.83 & 15.784 & 15.754 & 1.13 & 0.59 & 15.309 & 15.298 & 0.75 & 0.32 & Bl \\ 
V122        & 0.49762  & & & & & & & & & \\
V123 OoI   & 0.54547  & 16.043 & 15.952 & 1.404 &  & 15.700 & 15.648 & 1.113 & &  15.195 & 15.174 & 0.734 & & \\
V124 OoII  & 0.75243  & 15.987 & 15.977 & 0.449 &  & 15.579 & 15.573 & 0.335 & &  15.004 & 15.001 & 0.233 & & \\
V130 OoI   & 0.56899  & 16.049 & 16.035 & 1.14 & 0.26 & 15.614 & 15.607 & 0.78 & 0.17 & 15.088 & 15.085 & 0.53 & 0.12 & Bl \\ 
V133 OoI   & 0.55073  & 16.107 & 16.032 & 1.45 & 1.10 & 15.779 & 15.736 & 1.16 & 0.92 & 15.221 & 15.208 & 0.64 & 0.54 & Bl  \\ 
V134        & 0.61805  & & & & & & & & & Bl\\
V135 OoI   & 0.56839  & 15.962 & 15.918 & 1.17 & 0.80 & 15.630 & 15.605 & 0.91 & 0.58 & 15.181 & 15.171 & 0.60 & 0.36 & Bl  \\ 
V136        & 0.61720  & & & & & & & & & \\
V137 OoI   & 0.57515  & 16.066 & 16.003 & 1.193 &  & 15.720 & 15.686 & 0.921 & &  15.185 & 15.172 & 0.598 & & \\
V139        & 0.55999  & & & & & & & & & \\
V142 OoI   & 0.56863  & 16.120 & 16.047 & 1.297 &  & 15.763 & 15.723 & 1.003 & &  15.318 & 15.301 & 0.704 & & \\
V143 OoII  & 0.59641  & 15.846 & 15.772 & 1.54 & 1.20 & 15.495 & 15.458 & 1.16 & 0.90 & 15.035 & 15.016 & 0.78 & 0.60 & Bl  \\ 
V144$^{***}$ OoI   & 0.59670  & 16.043 & 16.024 & 0.60 & 0.60 & 15.649 & 15.641 & 0.42 (0.74) & 0.42 & 15.115 & 15.112 & 0.25 & 0.25 & Bl \\ 
V145        & 0.51449  & & & & & & & & & \\
V146 OoI   & 0.50219  & 16.022 & 15.926 & 1.402 &  & 15.677 & 15.628 & 1.046 & &  15.254 & 15.235 & 0.708 & & \\
V148        & 0.46728  &&&&& & & & & & & & \\
V149        & 0.54816  &&&&&&&&&&&&& Bl \\   
V150 OoI   & 0.52392  & 16.038 & 15.954 & 1.50 & 1.10 & 15.700 & 15.657 & 1.12 & 0.88 & 15.220 & 15.205 & 0.72 & 0.54 & Bl  \\ 
V151        & 0.51682  &&&&&&&&&&&&& Bl \\
V156        & 0.53200  & & & & & & & & & \\         
V157        & 0.54285  &&&&&&&&&&&&& Bl  \\            
V159        & 0.53382  &&&&&&&&&&&&& Bl \\  
V160        & 0.65730  &&&&&&&&&&&&& Bl \\       
V161        & 0.52656  &&&&&&&&&&&&& Bl \\ 
V165        & 0.48363  &&&&&&&&&&&&& Bl \\          
\end{tabular}
\end{center}
\end{minipage}
\end{table*}

\begin{table*}
\begin{minipage}{\textwidth}
\begin{center}
\contcaption{}
\begin{tabular}{l@{\hspace{2mm}}l@{\hspace{3mm}}c@{\hspace{1mm}}c@{\hspace{2mm}}l@{\hspace{1mm}}l@{\hspace{3mm}}c@{\hspace{1mm}}c@{\hspace{2mm}}l@{\hspace{1mm}}l@{\hspace{3mm}}c@{\hspace{1mm}}c@{\hspace{2mm}}l@{\hspace{1mm}}l@{\hspace{8mm}}l}
\hline
Star  &Period &$<B>^{a}$ &($B$)$^a$&\multicolumn{2}{c}{$A_B^b$} &$<V>$&($V$)&\multicolumn{2}{c}{\hskip 6mm $A_V^{b,c}$} &$<I_{\mathrm C}>$& ($I_{\mathrm C}$)      &\multicolumn{2}{c}{${A_{I_{\mathrm C}}}^b$} & Comment\\
 & \multicolumn{1}{c}{d}&mag&mag&mag&mag&mag&mag&mag&mag&mag&mag&mag&mag&\multicolumn{1}{l}{}  \\ 
\hline
V167 OoI & 0.64395  & 16.063 & 16.051 & 0.55 & 0.43 & 15.659 & 15.653 & 0.43 & 0.31 & 15.127 & 15.122 & 0.27 & 0.21 & Bl \\ 
V172      & 0.54229  & & & & & & & & & \\
V173      & 0.60700  & & & & & & & & & \\
V174      & 0.59123  &&&&&&&&&&&&& Bl  \\ 
V175      & 0.56970  & & & & & & & & & \\        
V176      & 0.53959  &&&&&&&&&&&&& Bl \\ 
V180      & 0.60910  & & & & & & & & & \\
V181      & 0.66380  & & & & & & & & & \\
V184      & 0.53128  &&&&&&&&&&&&& Bl \\       
V186      & 0.66327  &&&&&&&&&&&&& Bl \\
V187      & 0.58620  & & & & & & & & & \\
V189      & 0.61292  & & & & & & & & & \\
V190      & 0.52280  & & & & & & & & & \\
V191      & 0.51918  &&&&&&&&&&&&& Bl \\     
V192      & 0.49727  &&&&&&&&&&&&& Bl \\
V193      & 0.74786  & & & & & & & & & \\
V194      & 0.48920  & & & & & & & & & \\          
V195      & 0.64397  &&&&&&&&&&&&& Bl \\          
V197      & 0.49995  & & & & & & & & & \\
V201      & 0.54052  &&&&&&&&&&&&& Bl \\ 
V202 OoII &0.77359  & 15.962 & 15.960 & 0.191 &  & 15.557 & 15.556 & 0.146 & &  14.983 & 14.983 & 0.096 & & \\
V205 OoI  &0.63691  & 16.105 & 16.085 & 0.627 &  & 15.706 & 15.695 & 0.474 & &  15.154 & 15.149 & 0.322 & & \\
V211      & 0.55820  &&&&&&&&&&&&& Bl \\ 
V212      & 0.54222  &&&&&&&&&&&&& Bl \\
V214      & 0.53953  &&&&&&&&&&&&& Bl \\
V215      & 0.52863  & & & & & & & & & \\ 
V218 OoI & 0.54487  & 16.075 & 16.013 & 1.29 & 0.90 & 15.706 & 15.674 & 0.98 & 0.68 & 15.203 & 15.190 & 0.66 & 0.40  & Bl \\ 
V219      & 0.61363  &&&&&&&&&&&&& Bl \\ 
V220      & 0.60016  &&&&&&&&&&&&& Bl\\ 
V222 OoI & 0.59674  & 16.081 & 16.027 & 1.106 &  & 15.668 & 15.641 & 0.813 & &  15.168 & 15.157 & 0.550 & & \\
V226      & 0.48844  & & & & & & & & & \\
V229      & 0.51608  & & & & & & & & & \\
V234      & 0.50806  & & & & & & & & & \\
V235 OoII & 0.75985 & 15.853 & 15.834 & 0.632 &  & 15.482 & 15.471 & 0.495 & &  14.877 & 14.873 & 0.328 & & \\
V239      & 0.50399  &&&&&&&&&&&&& Bl\\
V241      & 0.59615  & & & & & & & & & \\ 
V243      & 0.63462  &&&&&&&&&&&&& Bl\\
V244      & 0.53785  & & & & & & & & & \\ 
V247      & 0.60535  & & & & & & & & & \\
V248      & 0.50980  & & & & & & & & & \\
V249      & 0.53300  & & & & & & & & & \\
V250a     & 0.5744   & & & & & & & & & \\
V250b     & 0.5924   & & & & & & & & & \\
V254      & 0.60563  & & & & & & & & & \\
V255      & 0.57265  &&&&&&&&&&&&& Bl \\
V257      & 0.60205  & & & & & & & & & \\
V258      & 0.71340  & & & & & & & & & \\
V262      & 0.65508  & & & & & & & & & \\
V270a     & 0.69018  & & & & & & & & & \\
V270b     & 0.62585  &&&&&&&&&&&&& Bl \\    
V271      & 0.63280  & & & & & & & & & \\
\multicolumn{11}{l}{New Var}\\
N.V.      & 0.24924  & 15.897 & 15.897 & 0.066 &  & 15.675 & 15.675 & 0.051 & &  15.453 & 15.452 & 0.033 & & \\          
\hline
\multicolumn{15}{l}{$^a$ $<>$ and () denote magnitude and intensity averaged mean magnitudes, respectively.}  \\ 
\multicolumn{15}{l}{$^b$ Peak to peak amplitudes; the observed largest- and smallest-amplitudes of Blazhko stars.}\\
\multicolumn{15}{l}{$^c$ Larger/smaller extreme value of the peak to peak amplitude  according to other data is given in parenthesis.}\\
\multicolumn{15}{l}{$^*$ $f_1$ appears in the residual with 0.005 mag amplitude in the $V$ band ($f_1=2.5799$ d$^{-1}$; $P_1=.3876$ d; $P_1/P_0=0.744$).}\\ 
\multicolumn{11}{l}{$^{**}$ Incomplete phase coverage, mean magnitudes and amplitudes cannot be determined. }\\ 
\multicolumn{11}{l}{$^{***}$ Amplitude modulation on a very long time-scale, see fig.~2. in \cite{rrl}. }
\end{tabular} 
\end{center} 
\end{minipage}
\end{table*}

The length and the time-coverage of the photometric data made it possible to determine the light-curve solutions for most of the stars including the Blazhko-modulated ones. We have found that the light curve of about  half (47\,percent) of the measured RRab stars show the Blazhko effect.  The true percentage of Blazhko stars may be even larger, as the photometry of the most crowded stars is too noisy to detect possible modulation.

The basic photometric results of stable-light-curve and Blazhko RRab stars are summarised in Table~\ref{rrab.dat}. The ID, the Oosterhoff type (based on the mean magnitudes and amplitudes as documented in Fig.~\ref{plotab}), the pulsation period, the magnitude and intensity averaged $BVI_{\mathrm {C}}$  mean magnitudes, and the amplitudes -- the smallest and largest amplitudes according to our observations for Blazhko stars -- are listed. If larger or smaller $V$ amplitude of a Blazhko star has been detected  in any of the published CCD data [\citet{K98}, \citet{co01},  \citet{H05}, \citet{be06}, and \citet{oc}], this extreme value of the amplitude is given in parenthesis. 

 Comparing the sample of the OoII stars with the results of \cite{ca05} a good agreement is found. Each of the stable-light-curve RRab stars classified as over-luminous by \cite{ca05} are OoII-type variable according to our photometry and vice versa. The Oosterhoff classification of some Blazhko stars (V035, V048, V067, V078 and V143) are, however, different, probably because the time coverage of the \cite{co01} observations does not make it possible to determine the mean magnitudes accurately enough if the amplitude of the light curve varies.

 A more detailed analysis and comparison of the photometric and spectroscopic results of Blazhko and stable RRab stars is planned to be published in a separate paper.

\begin{table} 
\begin{center} 
\caption{Summary of the radial-velocity data of overtone and double-mode RR Lyrae stars in M3.\label{rvc.sum}} 
\begin{tabular}{l@{\hspace{1mm}}c@{\hspace{1mm}}c@{\hspace{3mm}}c@{\hspace{3mm}}c@{\hspace{3mm}}l@{\hspace{3mm}}} 
\hline
Star & \multicolumn{2}{c}{Observation}& $\gamma$  &$A_{\mathrm{RV}}$&Comment$^{a}$  \\ 
& Hydra&Hect.&[km\,s$^{-1}$]& [km\,s$^{-1}$]&\\ 
\hline
V013 &    x & - & $-$139.0& 35.0& dm $f_{0.61}$\\
V037 &    x & - & $-$143.2& 26.6&    $f_{0.61}$ \\    
V056 &    x & x & $-$143.4& 26.2&    $f_{0.61}$\\
V068 &    x & - & $-$148.7& 23.5& dm $f_{0.61}$\\ 
V086 &    x & x & $-$146.4& 27.4&\\
V087 &    - & x & $-$149.6& 25.0& dm $f_{0.61}$\\
V088 &    x & - & $-$150.5& 10.8& \\
V097 &    x & x & $-$151.1& 24.0&    $f_{0.61}$\\
V107 &    x & x & $-$145.7& 29.5&\\
V125 &    x & - & $-$145.8& 22.9& dm $f_{0.61}$\\
V140$^{b}$&-& x & $-$160.0& 30.0&  Bl     \\
V203 &    x & - & $-$151.1& 10.4&  Bl  \\
New Var.& - & x & $-$143.8&  2.7&\\
\hline
\multicolumn{6}{l}{$^{a}$ Additional features of the light curve; see in \cite{overtone}. }\\
\multicolumn{6}{l}{$^{a}$ Data are derived from the entire Hectochelle observations. }
\end{tabular} 
\end{center} 
\end{table} 

\begin{table} 
\begin{center} 
\caption{Summary of the radial-velocity data of stable-light-curve RRab stars in M3.\label{rvab.sum}} 
\begin{tabular}{l@{\hspace{3mm}}c@{\hspace{1mm}}c@{\hspace{3mm}}c@{\hspace{3mm}}c@{\hspace{3mm}}}
\hline
Star$^a$ & \multicolumn{2}{c}{Observation}& $\gamma$  & $A_{\mathrm{RV}}$ \\ 
& Hydra&Hect.&[km\,s$^{-1}$]& [km\,s$^{-1}$]\\ 
\hline
V009  &  x & x & $-$152.6& 63.3\\
V015  &  x & - & $-$149.0& 61.7\\
V018  &  x & x & $-$140.2& 62.8\\
V019  &  - & x & $-$148.9& 40.9\\
V021  &  - & x & $-$149.2:& 62.0:\\
V026  &  - & x & $-$153.8& 60.8\\
V027  &  - & x & $-$152.2& 59.1\\
V036  &  x & x & $-$139.0& 63.6\\
V040  &  - & x & $-$150.2& 62.4\\
V046  &  x & x & $-$143.0& 45.6\\
V051  &  x & x & $-$147.4& 57.0\\
V055  &  - & x & $-$150.6& 62.7\\
V060  &  x & x & $-$145.3& 51.5\\
V064  &  - & x & $-$148.5:& 55.6:\\
V072  &  - & x & $-$145.2& 64.2\\
V081  &  x & x & $-$141.5& 63.1\\
V082  &  x & x & $-$147.0& 62.8\\
V083  &  - & x & $-$143.6& 63.1\\
V084  &  x & - & $-$143.6& 55.4\\
V089  &  - & x & $-$151.0& 63.7\\
V090  &  x & - & $-$140.1& 67.8\\
V093  &  x & x & $-$143.8& 55.8\\
V094  &  x & x & $-$146.7& 62.9\\
V096  &  - & x & $-$146.7& 65.5\\
V113  &  x & x & $-$145.4& 63.0\\
V115$^*$ &  x & x & $-$155.4& 65.7\\
V116  &  x & x & $-$143.6:& 64.5:\\
V120  &  - & x & $-$139.7& 39.8\\
V123$^*$  &  x & x & $-$147.9& 66.2\\
V124  &  - & x & $-$141.2& 31.8\\
V202  &  x & - & $-$148.5& 15.2\\
V206  &  x & - & $-$148.6:& 59.0:\\
\hline
\multicolumn{5}{l}{$^{a}$ * denotes stars with erroneous positioning. }
\end{tabular} 
\end{center} 
\end{table} 

\begin{table} 
\begin{center} 
\caption{Summary of the radial-velocity data of Blazhko RRab stars in M3.\label{rvbl.sum}} 
\begin{tabular}{l@{\hspace{0mm}}c@{\hspace{-0mm}}c@{\hspace{0mm}}l@{\hspace{0mm}}l@{\hspace{0mm}}l@{\hspace{2mm}}l@{\hspace{1mm}}l@{\hspace{1mm}}l}
\hline
Star & \multicolumn{2}{c}{Instr.}& \multicolumn{3}{c}{$\gamma^{*}$}  & \multicolumn{3}{c}{$A_{\mathrm{RV}}^{*}$.}
 \\ 
& Hyd.&Hect.&\multicolumn{3}{c}{[km\,s$^{-1}$]}& \multicolumn{3}{c}{[km\,s$^{-1}$]}\\ 

&&&\multicolumn{1}{c}{all}&Hect1 & Hect2&\multicolumn{1}{c}{all}&Hect1& Hect2\\
\hline
V010        &  x & - & $-$148.3 &&         & 60.6 &&\\ 
V014        &  - & x && $-$153.7 &         && 63.6 &    \\ 
V020        &  x & x & $-$151.0:&&         & 48.0:&&\\
V024        &  - & x && $-$139.0 & $-$138.6&& 53.4 & 60.0\\
V033        &  - & x &&          & $-$148.2&&      & 51.7\\ 
V034        &  - & x && $-$141.3 & $-$141.8&& 36.0 & 37.4\\ 
V035        &  - & x && $-$145.5 & $-$144.0&& 59.6 & 27.8\\ 
V038        &  - & x && $-$154.9 &         && 69.2 &     \\
V039$^{a}$  &  x & x & $-$140.0 &&         & 40.9:&&     \\ 
V043        &  - & x & $-$156.3 &&         & 56.0 &&\\ 
V045        &  x & x &&  & $-$153.4&& & 58.9\\ 
V048        &  - & x && $-$141.4 &         && 45.2 &     \\
V050        &  - & x & $-$145:  &&         & $>50$&&     \\
V052        &  x & x && $-$141.6 &         && 75.0 &     \\
V059        &  x & - & $-$150.5 &&         & 55.0 &&\\
V061        &  x & x && $-$148:  &         && 46.0:&     \\
V062$^{a}$  &  x & x & $-$143.5 &&         & 47.0 &&     \\
V063        &  x & x && $-$147.8 & $-$147.7&& 52.7 & 58.6\\ 
V066        &  x & x && $-$147.0 & $-$149.4&& 43.7 & 54.1\\ 
V067        &  - & x && $-$144.0 & $-$143.3:&& 52.5 & 70.5:\\ 
V071        &  - & x && $-$143.9 & $-$143.6&& 53.2 & 60.0\\ 
V073        &  - & x && $-$143.2 & $-$143.4&& 26.4 & 24.0\\ 
V077$^{a}$  &  x & - & $-$152.4 &&         & 57.0 &&     \\
V078        &  x & x && $-$142.4 & $-$142.6&& 65.4 & 53.4\\ 
V091        &  - & x & $-$145:  &&          & $>55$&&  \\
V104        &  - & x && $-$143.7  & $-$143.1&& 63.4 & 61.1 \\ 
V106        &  - & x &&  & $-$146.9&&  & 53.5 \\ 
V110        &  - & x &&           & $-$153.1&&      & 67.1\\ 
V114$^{a}$  &  x & x & $-$147.8  &&         & 58.4 &&\\
V117        &  - & x && $-$147.0: &         && 50.2:   &     \\
V130$^{b}$         &  - & x && $-$150.8  & $-$154.5:&& 27.3 & 52.3\\
            &    &   && $-$       & $-$152.8:&&      & 58.2\\
V144$^{c}$  &  x & - & $-$143.1  &&         & 2    &&\\
V161        &  x & - & $-$142.7  &&         & 23    &&\\
V186        &  - & x & $-$146.9  &&         & 38.0 &&\\
\hline
\multicolumn{9}{l}{$^{*}$ all: according all the RV data;}\\ 
\multicolumn{9}{l}{$^{*}$ Hect1: the first run of the Hectochelle observations.}\\
\multicolumn{9}{l}{$^{*}$ Hect2: the second run of the Hectochelle observations.}\\
\multicolumn{9}{l}{$^{a}$ Data are derived from the entire Hectochelle observations. }\\
\multicolumn{9}{l}{$^{b}$ Two solutions are given for the scarce large-amplitude phase.}\\
\multicolumn{9}{l}{$^{c}$ Large-amplitude modulation on a very long timescale.}
\end{tabular}
\end{center} 
\end{table}

\subsection{Radial-velocity results}\label{rv-result}
The RV data of overtone, non-Blazhko, and Blazhko RRab stars phased with the pulsation period are shown in Figs.~\ref{rvc}, \ref{rvab}, and \ref{rvbl}, respectively. Both the  Hectochelle@MMT and the Hydra@WIYN data are plotted. The Hectochelle@MMT RV data cover fully or partially the pulsation cycles at two epochs separated by about 30 days. The RVs of these two observation runs are shown by different symbols for Blazhko stars.

The $\gamma$-velocities and RV amplitudes ($A_{\mathrm{RV}}$) of overtone, single-mode and Blazhko RRab stars are listed in Table~\ref{rvc.sum}, \ref{rvab.sum}, and \ref{rvbl.sum}, respectively. The 2nd and 3rd columns of these tables identify the RV observations of the star. Additional features  of the light variations of the overtone variables \citep[see in][]{overtone} are indicated in the last column of Table~\ref{rvc.sum}.
The $\gamma$-velocity and $A_{\mathrm{RV}}$ correspond to the $a_0$ term and to the peak to peak amplitude according to an appropriate-order Fourier fit to the RV time-series, respectively.
The Hectochelle@MMT RV data are about 10 times more accurate than the Hydra@WIYN results, therefore, 
the Hydra@WIYN data are taken into account in deriving the $\gamma$-velocities and the RV amplitudes only for stars with no or incomplete phase-coverage by the Hectochelle@MMT data.

When it is possible, the mean and the amplitude of the RV curves are listed separately for the two runs of the Hectochelle@MMT observations of Blazhko stars. Columns $4-6$ and $7-9$  in Table~\ref{rvbl.sum} list the $\gamma$-velocity and $A_{\mathrm {RV}}$ derived from different parts of the observations, separately. See also the footnotes of Table~\ref{rvbl.sum} for further details. 

The rms scatter of the RV curves of single-mode RRLs in the Hectochelle@MMT data is typically $\sim0.5-0.7$\,km\,s$^{-1}$, i.e.,  apart from the systematic errors, the mean and the amplitude of the RV curves are determined with about 0.1 and 0.3\,km\,s$^{-1}$ accuracy, respectively, if the phase coverage is complete.  16 of the 79 variables observed spectroscopically have only Hydra@WIYN measurements. Taking into account the $5-7$\,km\,s$^{-1}$ uncertainty of these RV data, and the number of observations available for the variables ($60-80$), the mean RV values are determined with $0.5-0.8$\,km\,s$^{-1}$ accuracy for these stars. 

The uncertainties of the mean and the amplitude values for variables with improper phase coverage are larger than estimated for the other stars. 

There are two stars, V115 and V123, for which positioning was not accurate enough. Consequently, the Hectochelle@MMT spectra of these stars have smaller S/N ratio than for the other stars and the rms scatter of their derived RV curves is $2-3$\,km\,s$^{-1}$. 

The differences between the mean RVs  determined at the different-amplitude phases of the modulation of at least 4 Blazhko stars  (V035, V066, V104 and V130) seem to be larger than that the  uncertainties would indicate. The  absolute value of the mean RV is higher at the large-amplitude phase than at the small-amplitude one in each of these cases. We do not think, however, that the differences in the mean RVs of Blazhko stars at different phases of the modulation would indeed mean real changes of the $\gamma$-velocity of the stars. Instead, most probably, the dynamics of the atmosphere produce these apparent discrepancies, at least partly. The RV is determined from metallic lines, which are formed at different depth in different phases of the pulsation. However, the motion of the atmosphere is not homogeneous, and strong velocity gradients appear not only during the most violent phases of the pulsation \citep{cp, gf}.  As a consequence, the mean of the derived RVs does not necessarily reflect the real $\gamma$-velocity of the star. 

Moreover, despite the strong line doubling, which characterises the spectra during the brightening phase of the pulsation, making these RVs doubtful \citep[read also][for a detailed discussion on the atmospheric dynamics of RRL stars]{preston}, these RVs are also counted when calculating the mean. As these effects may be amplitude dependent, they can explain the apparent connection between the amplitude and the sign of the bias in the mean RVs of Blazhko stars. 

Another source of an amplitude-dependent bias on the RVs and on their mean is in the selection of the synthetic template spectrum for the reduction process (described in Sect.~\ref{sect.hecto}). It was found that the amplitude of the deduced RV curve depends, in some extent, on the temperature of the template used. The amplitude of the $T_{\mathrm{eff}}$ variation during the pulsation of RRL stars can be as large as 2000 K. Therefore, using a fixed-temperature template to derive the CCF of the spectra, an amplitude-dependent bias of the results is generated.

\subsection{A new overtone variable}\label{sect.nv}

\begin{figure}
\centering
\includegraphics[width=7.9cm]{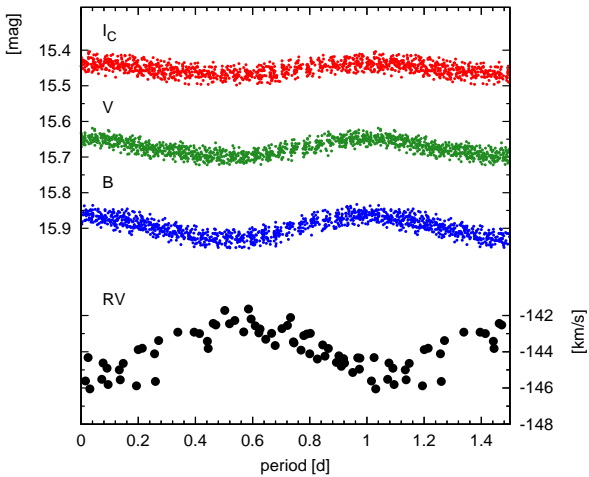}
\caption{ Phased $BVI_{\mathrm{C}}$ light curves and RV curve of the new variable.\label{ujv}}
\end{figure}

Spectra of non-variable stars were also observed with the Hectochelle@MMT for calibration purposes. The RV data of one of these stars turned out to  show very small-amplitude, short-period RV variation. The new variable is at a large, 24.5 arcmin  radial distance from the cluster centre, at the position of 13:41:18.85 +28:01:57 (J2000).  The magnitudes show  periodic variation with a $\sim0.25$\,d periodicity, in accordance with the RV variation. The phased $BVI_{\mathrm{C}}$ light curves and the RV data of the new variable are shown in Fig.~\ref{ujv}.
The mean $B,V$ and $I_\mathrm{C}$ magnitudes (15.90, 15.68 and 15.45 mag, respectively) of the star fit the blue edge of the instability strip of overtone variables on the horizontal branch  \citep[cf. Fig. 3 in][]{overtone}. Its very small amplitude, short period, and blue colour indicate that the new variable is a second-overtone RRL belonging to M3. Its $\gamma$-velocity ($-143.8$\,km\,s$^{-1}$) and proper motion  \citep[$\mu_{\alpha}{\mathrm{cos}}\delta=-1.9$ mas\ yr$^{-1}$,  $\mu_{\delta}=-0.5$ mas\ yr$^{-1}$][]{smart}  agree with the mean cluster velocity  ($-147$\,km\,s$^{-1}$, see in Sect.~\ref{sect.vrad}) and absolute proper motion \citep[$\mu_{\alpha}{\mathrm{cos}}\delta,\mu_{\delta} =-2.5\pm0.6, -2.1\pm0.8$ mas\ yr$^{-1}$,][]{tuch} reasonably well.

\section{Fourier parameters of the $V$ light curve and the radial-velocity curve of stable RRab stars}

\begin{figure*}
\centering
  \centering
  \includegraphics[width=8.7cm]{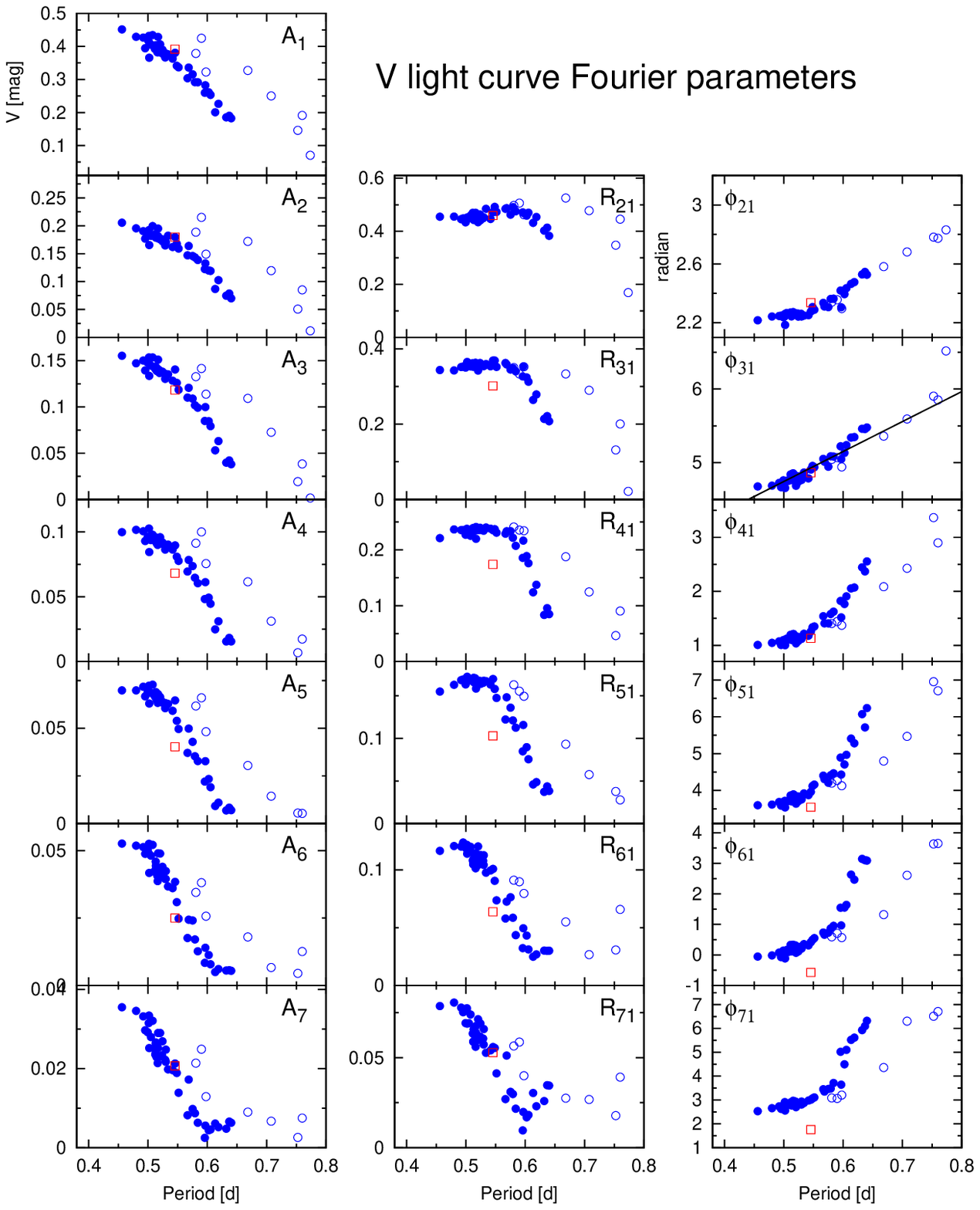}
  \centering
  \includegraphics[width=8.7cm]{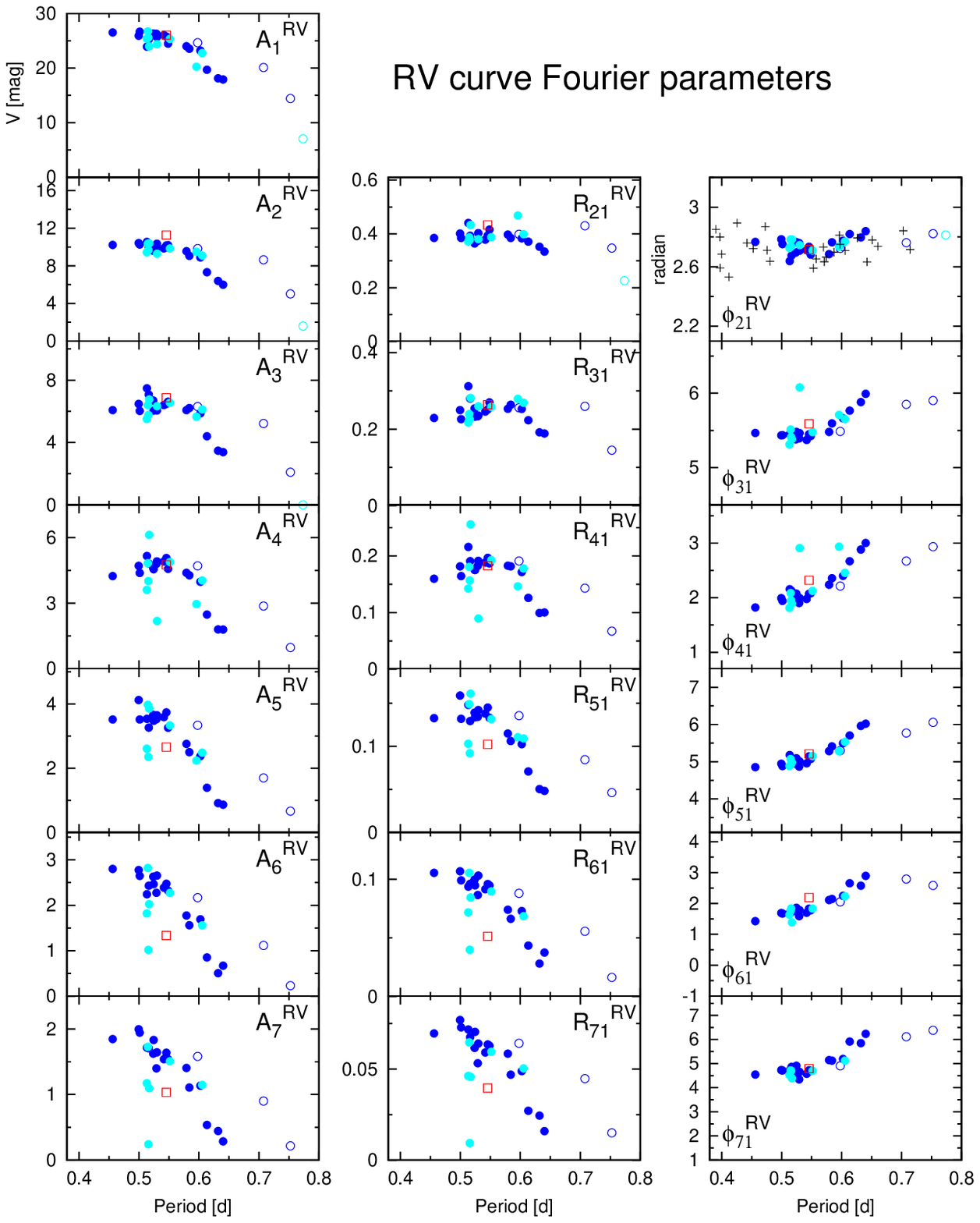}
\caption{The Fourier parameters of the V light curves and the radial-velocity curves of stable RRab stars in M3 are shown in the left- and right-hand panels, respectively. The Fourier amplitudes, amplitude ratios, and epoch-independent phase differences of $sin$ decomposition  ($A_1-A_7, R_{21}-R_{71}$ and $\Phi_{21}-\Phi_{71}$) are plotted. Open symbols denote OoII variables. The light-blue symbols in the plots of the RV data denote uncertain values, variables with noisy and/or scarce RV data.  The ${\Phi_{21}}^{\mathrm RV}$ values  of Galactic field RRab stars are also shown for comparison (black crosses in the top/right-hand panel). Note that the scales of the $R_{k1}$ and ${R_{k1}}^{\mathrm RV}$ and the  ${\Phi_{k1}}$ and ${\Phi_{k1}}^{\mathrm RV}$  amplitude-ratio  and phase-difference panels are identical, respectively,  for a better comparison. The line drawn in the $Period-\Phi_{31}$ panel corresponds to [Fe/H]=$-1.45$ according to the metallicity formula derived in \citet{jk96}. Results for the peculiar V123 are shown by red square symbols.}
\label{four}
\end{figure*}

\begin{table} 
\begin{center} 
\caption{Comparison of the statistics of the ${\Phi_{21}}^{\mathrm RV}$ phase differences of the radial-velocity curves of Galactic field and M3 RRab stars. \label{phi21}} 
\begin{tabular}{cccccc}
\hline
      &    &        &  ${\Phi_{21}}^{\mathrm RV}$&      &        \\
Sample& N  &  min.  & max.      &  mean   & s.d.\\
\hline
M3& 31&2.64&2.84&2.74&0.05\\
field& 26&2.53&2.89&2.72&0.09\\
\hline
\end{tabular} 
\end{center}
\end{table}

The Fourier parameters of the light curves of RRL stars are good indicators of the physical properties of the stars \citep{st,jk96,j98,kw}, but systematic studies of the Fourier parameters of the RV curves are available only  for Cepheids \citep{m00,pont}.
\cite{liu} has found that, in contrast with the large diversity of the shapes of the light curves,  the shapes of the available sample of RRL RV curves is relatively stable; actually only their amplitudes are scaled according to the amplitude of the light curve.  

The Fourier parameters of the $V$ light-curves and the RV curves of stable RRab stars in M3 are compared in Fig.~\ref{four}. The light-curve parameters follow the known, period dependent trends, and the positions of the OoI and OoII stars are well separated in the panels showing the amplitudes, amplitude ratios and the higher-order phase differences. In contrast, the OoI and OoII variables follow the same tracks in the panels showing the lower-order phase differences. Among the Fourier parameters of the light curve shown, the $\Phi_{31}$ data define the most linear relation with the period, in accordance with the metallicity formula \citep[][JK96]{jk96}, which predicts a linear relation between these parameters for a sample of different period variables of the same metallicity.
The straight line in this panel corresponds to the JK96 relation for [Fe/H]=$-1.45$. The only star, which deviates from this relation significantly, is the longest period, smallest amplitude, OoII star, V202. The light curve of this star is sinusoidal, consequently  its 3rd-order Fourier parameters are uncertain.

The amplitudes and amplitude ratios of the RV curves follow the same trends as the Fourier parameters of the light curve. Outliers are only among the poor-quality RV data shown by light-blue symbols in Fig.~\ref{four}. What is, however, indeed surprising, is the behaviour of the phase differences of the RV curves. The full range of the variation of the $\Phi_{21}$ ... $\Phi_{71}$ phase differences increases from 0.65 to 4.45 for the light-curve parameters, but only from 0.20 to 2.03 for the  ${\Phi_{k1}}^{\mathrm RV}$ parameters. (Note that, for a better comparison, the scales of the amplitude-ratio and phase-difference plots for the $V$ and RV data are identical.) 

Considering ${\Phi_{21}}^{\mathrm RV}$, no period dependence of this parameter is detected at all. 
To check the stability of the  ${\Phi_{21}}^{\mathrm RV}$ phase differences of the RV curves of RRab star, the data of Galactic field stars compiled by \cite{k03} are also shown in this panel by crosses. The [Fe/H]s of these field stars cover a wide metallicity range: from solar (0.0) to very metal poor ($-1.9$). The [Fe/H] of each field star with a period shorter than 0.5 d is larger than $-1.0$. In spite of the significant differences of the field and M3 samples, their  ${\Phi_{21}}^{\mathrm RV}$ phase differences seem to have very similar values. The statistics of the  ${\Phi_{21}}^{\mathrm RV}$ phase differences of the two samples are summarised in Table~\ref{phi21}. 

We thus conclude that the ${\Phi_{21}}^{\mathrm RV}$  phase difference of the RV curve of RRab stars is  uniformly constant, it does not show any detectable period or metallicity dependence larger than the uncertainty of the data.

According to $Z=0.001$, $M=0.65 M_\odot$ and $L=52.5 L_\odot$ convective, nonlinear pulsation models  with periods between $0.48 -0.80$~d \citep{feu} the  ${\Phi_{21}}^{\mathrm RV}$\,...\,${\Phi_{51}}^{\mathrm RV}$  phase differences of model RV curves vary within about twice as large ranges as observed in M3, from 0.5 (${\Phi_{21}}^{\mathrm RV}$) to 2.0 (${\Phi_{51}}^{\mathrm RV}$), and constancy of the ${\Phi_{51}}^{\mathrm RV}$ phase difference  is  predicted  in the $0.45-0.70$ d period interval, which does not seem to be valid based on the observations. 

\section{Radial dependence of the dispersion of the mean radial velocities of RR Lyrae stars}\label{sect.vrad}

\begin{table} 
\begin{center} 
\caption{Radial variation of the dispersion of the mean radial velocities of RR Lyrae stars and giants in M3. \label{rvdisp}} 
\begin{tabular}{lrrrcc}
\hline
RD$^{a}$  & N  &  mean RD & s.d. &  mean RV ($\gamma$-vel.)  & s.d.\\
arcmin    &    &  arcmin  &      &   km/s  &     \\
\hline
\multicolumn{6}{l}{RR Lyrae}\\
 0.0-2.5& 14 &  1.72 & 0.43 & -149.14 & 5.85 \\
 2.5-5.3& 32 &  3.71 & 0.87 & -145.95 & 4.75 \\
 5.3-7.5& 13 &  6.41 & 0.66 & -146.99 & 3.65 \\
 7.5-12.3& 16&  9.24 & 1.43 & -145.72 & 2.37 \\
12.3-   & 4  & 21.15 & 6.78 & -148.90  & 4.81 \\
\multicolumn{6}{l}{giants}\\
 0.0-2.5& 15 &  1.63 & 0.61 & -147.81 & 7.62 \\
 2.5-5.3& 26 &  3.71 & 0.76 & -147.22 & 4.88 \\
 5.3-7.5& 12 &  6.59 & 0.46 & -149.65 & 4.05 \\
 7.5-12.3& 48&  9.93 & 1.22 & -147.39 & 2.61 \\
12.3-   & 53 & 18.08 & 4.64 & -146.41 & 2.75 \\
\hline
\multicolumn{6}{l}{$^{a}$ Projected radial distance.}
\end{tabular} 
\end{center}
\end{table}

\begin{figure}
\centering
\includegraphics[width=7.6cm]{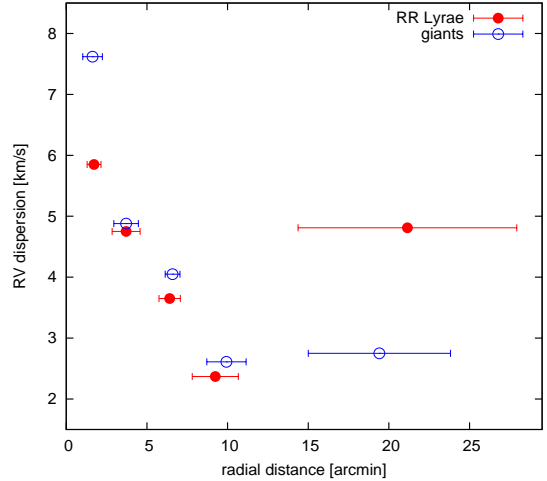}
\caption{Comparison of the projected radial dependence of the dispersion of the  $\gamma$-velocities of RR Lyrae stars and the radial velocities of giants \citep{kimmig} in M3. The points are shown at the mean value of the radial distances of the sub-samples and the horizontal bars denote their standard deviations.\label{raddist}}
\end{figure}

The statistics of the radial dependence of the dispersion of the RV data in GCs is a key function for dynamical studies. A large sample of RV data of M3 giants, observed also with the Hechtochelle@MMT instrument, were published recently by \cite{kimmig}.  Although other RV studies of M3 are also available in the literature \citep[e.g.,][]{gg,sod,Pil2000,Smol2011,Kam2000}, these studies include smaller number of objects, and there is some overlap between the sample stars in some cases. Therefore, we selected exclusively the \cite{kimmig} data for the comparison of the RV dispersion values of RRL and of other  giant stars in M3. The typical accuracy of the RV values given by \cite{kimmig} is some tenths of \,km\,s$^{-1}$.

The mean value of the $\gamma$-velocities of the 79 RRL stars listed in Table~\ref{rvc.sum},~\ref{rvab.sum} and ~\ref{rvbl.sum} is $-146.8$\,km\,s$^{-1}$ with 4.52\,km\,s$^{-1}$ rms. If multiple entries of the mean RV of Blazhko stars are given in Table~\ref{rvbl.sum}, the average of these values are taken here. This result is in good agreement with the $-147.4$\,km\,s$^{-1}$ mean RV with 4.05\,km\,s$^{-1}$ rms of the 139 giants \citep{kimmig}. The minor, 0.5\,km\,s$^{-1}$ smaller RV dispersion value of the Kimmig's data may be explained by the overpopulation of the sample stars in the outer regions in this data-set. While 
there are similar number of  stars  in the inner regions (projected radial distances RD $<12$ arcmin) in the samples of  RRL stars and Kimmig's giants, the two samples contain quite different number of stars in the outer regions. Only a few RRL but about half of the total sample of the giants are located here and the RV dispersion is known to be decreasing outwards in GCs.

Other RV measurments of M3 stars are also in good agreement with the above results, e.g. \cite{gg} determined $-146.9$\,km\,s$^{-1}$ cluster-mean velocity for a sample of 111 stars, while \cite{sod} derived $-147.0$\,km\,s$^{-1}$ for 87 giants.

The radial dependence of the statistics of the  mean RVs of the RRL stars (Table~\ref{rvc.sum}, \ref{rvab.sum}, \ref{rvbl.sum}) and of the \cite{kimmig} sample are summarised in Table~\ref{rvdisp} and the results are shown in Fig.~\ref{raddist}. The range of the  projected RDs, the number of stars within the RD range (N), the mean and the rms of the RDs and the mean  and the rms of the mean RVs  within the given RD bin are listed in the $1-6$ columns of Table~\ref{rvdisp}, respectively. The accepted binning was chosen to obtain the possible most similar N and mean RD values for the four inner bins of the two samples.

The RV dispersion of RRL stars is smaller than the RV dispersion of giants in the four inner bins as can be seen in Fig.~\ref{raddist}. However, the outermost bin contains only 4 RRL stars but 53 giants; thus the dispersions of the mean RVs of the two samples cannot be compared reliably here. As the mean RV can be determined less accurately for pulsating variables than for constant stars (see details in Sect.~\ref{spect} and \ref{rv-result}) this is just the opposite of what one would expect. Therefore, the statistical significance of the differences has to be checked to decide the reliability of this result. 

To test the significance of the apparent systematic differences between the two samples we use a simple, two-parameter fit to the data as defined in eq. 12. of \cite{hernandez} instead of a physically motivated more model-specific solution \citep[e.g.,][]{king}. The individual errors of the RV values are not taken into account in the fitting process, because they cannot be determined correctly for RRL stars if their phase coverage is not complete and because of the detected amplitude-dependent bias of the mean RVs. However, this is not relevant for our purpose, since the error of the mean RVs are supposed to be still significantly smaller than the intrinsic RV dispersion, especially in the inner RD bins.

We made the null hypothesis that the effect is purely accidental. Supposing this null hypothesis is valid, we merged the RRL and giant star data into one sample and fitted the radial dependence of the velocity standard deviation using eq. 12. of \cite{hernandez} 
\begin{equation}
\sigma (r) = \sigma_1 exp(-\frac{r^2}{{r_\sigma}^2}) + \sigma_0
\end{equation}
\noindent  where $\sigma_1. \, r_\sigma$, and $\sigma_0$ are constants determined by least squares fitting. Applying this formula, we computed the standardized sum of squares of the velocity differences by
\begin{equation}
    Q^2 = n^{-1} \sum \limits_{i=1}^n \frac{(RV_i-RV_{\mathrm{mean}})^2}{\sigma(r_i)^2}
\end{equation}
\noindent for the RRL and the giant stars separately ($n$ is the sample size in the above formula). If the null hypothesis is valid, the ${Q_{\mathrm{RRL}}}^2/{Q_{\mathrm{giants}}}^2$ ratio of the RRL and giant stars should not differ significantly from unity. In our case the value of this ratio is  0.82. Assuming a valid null hypothesis the probability for getting this value purely accidentally is $ p=0.18$. Consequently, the apparent difference between the standard deviations of the RRL and giant stars is not, in fact, significant statistically.

\section{Analysis of the combined photometric and radial-velocity data of stable-light curve RR Lyrae stars in M3}\label{sect.anal}

\subsection{Parallel light and radial-velocity curves}\label{sect.sim}

The simultaneous photometric light/colour curves and RV data  used in the analysis are calculated from
appropriate order ($3-16$) harmonic fits to the observations. The photometric data of overtone and double-mode variables were published in~\cite{overtone}.

When the Hydra@WIYN RVs are used in combination with the photometry,  the phase-match of the data are guaranteed by harmonizing the phases of the 2012 light curves with the photometric observations obtained in $1998-1999$ \citep{be06,oc}.
\begin{figure}[t]
\centering
\includegraphics[width=8.2cm]{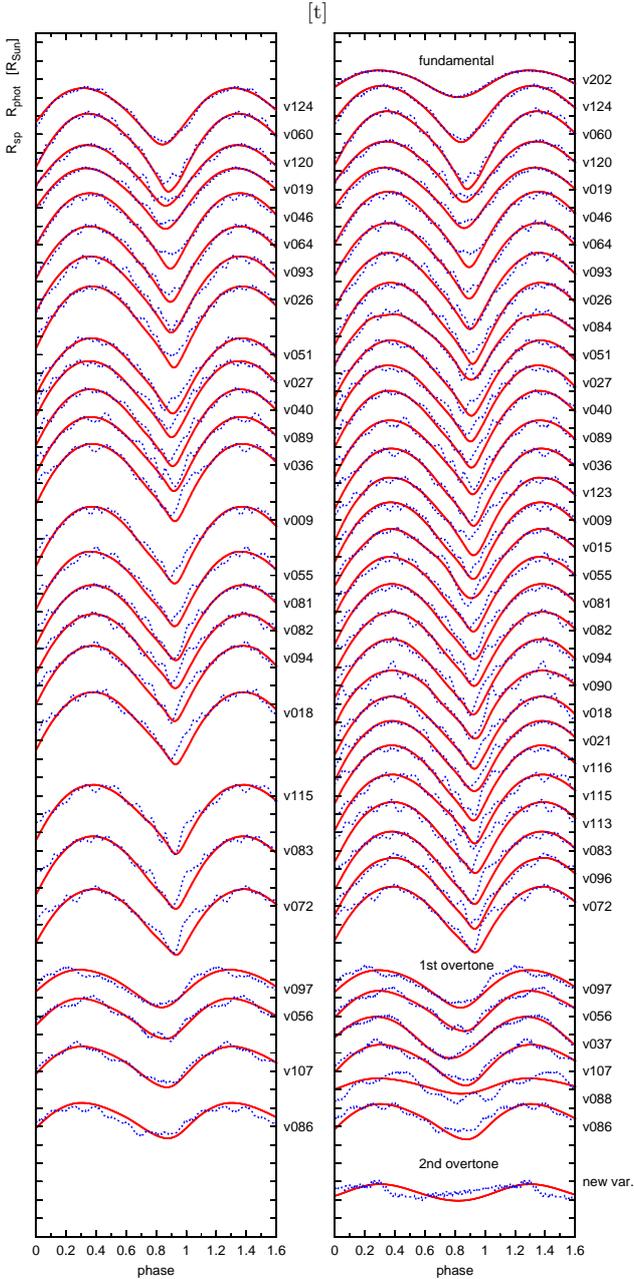}
\caption{Comparison of the photometric   ($R_{\textrm{phot}}$: blue dotted lines) and spectroscopic ($R_{\textrm{sp}}$: red continuous lines) radii variation of single-mode RR Lyrae stars in M3. The best-fit solutions with the distance as a free parameter are shown in the left-hand panel for variables with complete, high accuracy photometric and radial-velocity data.  Using a fixed, 10480~pc value for the distance of M3, the matches of the $R_{\textrm{phot}}$ and  $R_{\textrm{sp}}$ curves are shown in the right-hand panel. Stars are shown in the order of decreasing pulsation period from top to bottom.  The $R$ curves are vertically shifted for clarity, the tics on the $y$ axes denote 0.20, 0.10 and 0.02 $R_{\odot}$ for the fundamental, first- and second-overtone RR Lyrae stars, respectively.} 
\label{std.bw} 
\end{figure}

\subsection{Baade-Wesselink analysis}\label{sect.bw}

The RV and photometric data of M3 discussed in Sect.~\ref{sect.data} and published in \cite{overtone} provide the first possibility to perform classical BW analysis of a large sample of stars in a GC.

As the tidal radii of GCs are not significantly larger than 100 pc [\cite{n16} and \cite{h96} gave 113 and 85~pc values  for M3, respectively], the distances of the variables should have to be the same within about a  $100-200$~pc range in a given GC. Hence, the real star by star differences between the distances are expected to be smaller  than  the inherent $\sim5-10$\,percent uncertainty of the photometric BW method. Therefore, the analysis is performed in two steps: first, the distances of the variables with the best-accuracy and phase-coverage RV curves are determined, then, secondly, fixing the distance to the mean value of the distances of these stars, the mean radius and the radius variation are determined for the full sample of the variables observed both photometrically and spectroscopically. It was shown in \cite{overtone} that a small-amplitude signal appears in the spectra of many RRc stars at 0.61  frequency ratio. These frequencies have, however, mmag amplitudes only, therefore these variables are also included in the analysis.

Because of their high accuracy and  simultaneity with the photometry,  the Hectochelle  RV curves of the variables are  used, when available. The additional stars are 4 RRab (V015, V084, V090 and V202) and 2 RRc stars (V037 and V088) which have only Hydra data.  Some of the Hectochelle RV curves lack a complete phase coverage; to obtain a continuous fit for these data they are complemented with artificial points according to the Hydra data and/or taking into account the shape of the light curve. The uncertainty of this process has, however, a negligible effect on the distance determination of the cluster, and on  the obtained results.

\subsubsection{The method}
Supposing that the pulsation  is fully radial, the equality of the photometric and spectroscopic radii is the base idea of the BW method.   These radii are derived according to the following equations:
\begin{align}
R_{\textrm{ph}}&=d\times \theta,\,\, \theta=10^{0.2(-(V_0+BC)-10*\mathrm{log}(T_{\mathrm{eff}})+37.35)}\label{theta}\\
R_{\textrm{sp}}  &=R_0+p\times \int(V_{\textrm{rad}}-\gamma),\label{rad}
\end{align}
\noindent
where $R_{\textrm{ph}}$ and $R_{\textrm{sp}}$, $d$, $\theta$, $V_0$, B.C., $R_0$, $p$ and  $\gamma$ are the photometric and spectroscopic radii, the distance, the stellar angular radius, the $V$ magnitude corrected for interstellar absorption, the bolometric correction, the mean radius, the projection factor and the mean value of the RV, respectively. Note that, instead of AU and pc, the dimensions of $R$ and $d$  are $R_\odot$ and parsec in the first part of Eq.~\ref{theta}, thus $\theta$ is measured in 0.00456 arcsec unit. The B.C. scale is set to give $-0.07$ B.C.$_\odot$, and $M_{\mathrm{bol}\odot}=4.74$\,mag is regarded \citep{torres}. The value of the $p$-factor is taken to be 1.35 \citep{n04}. 
The effective temperature ($T_{\mathrm{eff}}$) and B.C. are determined from the dereddened ($V-I_{\mathrm{C}}$) colour and the effective $\mathrm{log}g$, using synthetic ($V-I_{\mathrm{C}}$) colours of static atmosphere models \citep{kurucz}. The effective $\mathrm{log}g$ is calculated as the sum of the static $\mathrm{log}g$ according to eq.~15. in \cite{j98}, which yields $\mathrm{log}g=2.7-2.9$ values for RRab stars, and the derivative of the RV curve. The static $\mathrm{log}g$ of the overtone variables is taken to be 3.0. The metallicity and reddening of M3 are adopted to be $-1.5$ and $E(B-V)=0.014$ mag, respectively \citep{h96}. The enhancement of the $\alpha$ elements is supposed to be 0.3 dex \citep{carney}.  The static atmosphere model grids are interpolated to obtain 0.05 and  5 K resolution in log$g$ and in $T_{\mathrm{eff}}$, respectively.

The distance and the mean radius are determined by matching the amplitudes and the zero points of the $R_{\textrm{ph}}$ and the $R_{\textrm{sp}}$ curves between pulsation phases 0.1/0.2 and 0.8/0.9 for fundamental mode and for RRc stars showing a pronounced bump on the light curve preceding maximum brightness and by using the complete $R$ curves  for overtone variables with sinusoidal light curves. For a detailed discussion on the effect of different pulsation phases on the results of BW analyses read e.g. \cite{ccf} and \cite{jcsl}.

\subsubsection{The first step: determination of the distance}
There are 22 RRab and 4 RRc stars with high S/N RV data of good phase coverage, which are appropriate  for the complete analysis. The  distances derived for these stars are listed in the third column of Table~\ref{bw}; the first and second columns list the ID of the star and the pulsation period, respectively. The $R_{\textrm{ph}}$ and $R_{\textrm{sp}}$ curves of the best fit for these 26 stars are shown in the left-hand panel of Fig.~\ref{std.bw}. The match between $R_{\textrm{ph}}$ and $R_{\textrm{sp}}$ is satisfactory; significant differences between the two curves appear only between phases 0.9 and 1.1, when the dynamics of the atmosphere is the most violent. 

BW distance moduli   were determined previously for only a very limited number of stars in the M4, M5, M92 and 47\,Tuc GCs with $0.15-0.20$ mag uncertainty  \citep{cohen,liu90,storm1,storm2}. The corresponding uncertainties of the distances are $5-10$\,percent. The BW distances of the 26 M3 stars cover the $9680-11420$~pc range with 10480~pc value of both the mean and the median. The standard deviation of the derived distances is 310 pc, i.e. about 3\,percent of the mean value.  

\begin{figure}
\centering
\includegraphics[width=8.2cm]{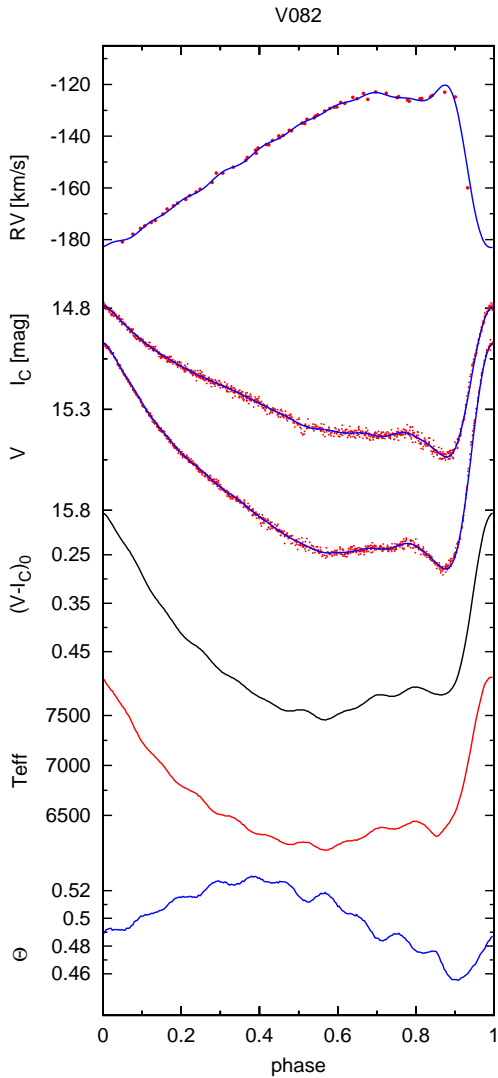}
\caption{The measured and fitted radial velocity, $V$ and $I_{\mathrm C}$ light curves, the $(V-I)_0$ colour curve calculated according to the Fourier fits of the light curves, the corresponding temperature, and  angular radius variation (in 0.00465 mas units) of V082 are shown in the figure.}
\label{v82} 
\end{figure}

The scatter of the distance estimates reflects  primarily the  photometric uncertainties and the uncertainty of the fitting process of the $R_{\textrm{ph}}$ and $R_{\textrm{sp}}$ curves.  The magnitude zero point (and amplitude) even of very-small-scatter light curves may be uncertain by $0.01-0.05$ mag in crowded-field photometry. A 0.02 mag error of the magnitude zero point of a given star results in about 300~pc difference of the derived distance. Even larger, about 500~pc difference in the distances may arise depending on that which part of the radii curves around the rising branch of the light curve is omitted from the fit. This is because of the inadequacy of both the photometric and spectroscopic data to derive the radius variation during the most violent phases of the pulsation and also because of the wiggles on the photometric radius curves. These wiggles originate form the uncertainty of the process:  the $V-I_{\mathrm{C}}$ colour curve is determined as the subtraction of the synthetic light curves derived as high order Fourier fits to the observations, and the marginal wiggles on the resultant $V-I_{\mathrm{C}}$ curve are magnified  when transforming the logarithmic scale to linear. For an example, the RV, $V$,$I_{\mathrm C}$,  $V-I_{\mathrm{C}}$, $T_{\mathrm{eff}}$ and angular radius variations of V082 are illustrated in Fig.~\ref{v82}.

\begin{table*} 
\begin{center} 
\caption{Results of the Baade-Wesselink analysis of single-mode RR Lyrae stars in M3. \label{bw}} 
\begin{tabular}{l@{\hspace{1mm}}l@{\hspace{6mm}}l@{\hspace{4mm}}rrrrrr}
\hline
Var.&Period & Distance&\multicolumn{6}{c}{ d=10480 pc} \\ 
&&&$R$& $\Delta R^a$&$T_{\mathrm{eff}}$&$\Delta T$&$T_{\mathrm{min1}}^b$&$T_{\mathrm{min2}}^c$\\
\hline
& days &~pc & \multicolumn{2}{c}{$R_\odot$}&\multicolumn{4}{c}{K}\\
\hline
 \multicolumn{9}{l}{Fundamental mode}\\
V009 & 0.541549  &  10600 &  5.33 & 0.84 & 6619 & 1625 & 6150 & 6425 \\
V015 & 0.530088  &  --    &  5.34 & 0.75 & 6647 & 1735 & 6185 & 6475 \\
V018 & 0.516455  &  10180 &  5.06 & 0.79 & 6681 & 1745 & 6235 & 6380 \\
V019 & 0.631980  &  10700 &  5.74 & 0.66 & 6377 &  670 & 6150 & 6255 \\
V021 & 0.515766  &  --    &  5.10 & 0.76 & 6678 & 1795 & 6205 & 6425 \\
V026 & 0.597751  &  10540 &  5.74 & 0.88 & 6527 & 1435 & 6125 & 6260 \\ 
V027 & 0.579063  &  10500 &  5.32 & 0.83 & 6631 & 1280 & 6280 & 6390 \\
V036 & 0.545596  &  10180 &  5.26 & 0.84 & 6678 & 1700 & 6210 & 6380  \\
V040 & 0.551539  &  10640 &  5.30 & 0.84 & 6585 & 1440 & 6185 & 6325 \\
V046 & 0.613388  &  10200 &  5.71 & 0.70 & 6394 &  805 & 6120 & 6290 \\
V051 & 0.583968  &  10500 &  5.49 & 0.82 & 6516 & 1235 & 6160 & 6320 \\
V055 & 0.529826  &  10320 &  5.32 & 0.81 & 6595 & 1755 & 6095 & 6360 \\
V060 & 0.707729  &  10280 &  6.00 & 0.85 & 6461 &  945 & 6165 & 6235 \\
V064 & 0.605461  &  10420 &  5.70 & 0.82 & 6415 & 1010 & 6110 & 6255  \\
V072 & 0.456079  &  10440 &  4.66 & 0.72 & 6895 & 2190 & 6325 & 6595 \\
V081 & 0.529120  &  11420 &  5.26 & 0.82 & 6622 & 1640 & 6170 & 6375 \\
V082 & 0.524539  &  10400 &  5.22 & 0.81 & 6647 & 1730 & 6150 & 6380 \\
V083 & 0.501270  &  10600 &  5.03 & 0.79 & 6750 & 1905 & 6230 & 6495 \\
V084 & 0.595729  &  --    &  5.72 & 0.71 & 6428 & 1080 & 6130 & 6195 \\
V089 & 0.548481  &  10680 &  5.39 & 0.81 & 6622 & 1540 & 6210 & 6285 \\
V090 & 0.517030  &  --    &  5.16 & 0.75 & 6658 & 1865 & 6165 & 6405 \\
V093 & 0.602297  &  10400 &  5.76 & 0.83 & 6420 & 1105 & 6080 & 6165 \\
V094 & 0.523696  &  10880 &  5.17 & 0.82 & 6660 & 1695 & 6215 & 6435 \\
V096 & 0.499406  &  --    &  5.21 & 0.77 & 6660 & 1965 & 6135 & 6295 \\
V113 & 0.513005  &  --    &  5.09 & 0.77 & 6721 & 1850 & 6220 & 6400 \\
V115 & 0.513346  &   9680 &  5.23 & 0.75 & 6650 & 1825 & 6130 & 6365 \\
V116 & 0.514809  &  --    &  5.11 & 0.82 & 6690 & 1760 & 6210 & 6440 \\
V120 & 0.640145  &  10380 &  5.89 & 0.66 & 6326 &  710 & 6070 & 6235 \\
V123 & 0.545475  &  --    &  5.42 & 0.85 & 6551 & 1695 & 6110 & 6270 \\
V124 & 0.752440  &  10660 &  6.34 & 0.61 & 6277 &  500 & 6090 & 6170 \\
V202 & 0.773575  &  --    &  6.38 & 0.29 & 6281 &  235 & 6180 & 6200 \\
 \multicolumn{9}{l}{1st overtone}\\
V037 & 0.326636  &  --    &  4.53 & 0.22 & 7175 &  740 & 6835 & 6895 \\
V056 & 0.329598  &  10180 &  4.69 & 0.22 & 7074 &  745 & 6740 & 6740 \\
V086 & 0.292656  &  10460 &  4.27 & 0.19 & 7348 &  880 & 6965 & 6975 \\
V088 & 0.298750  &  --    &  4.33 & 0.08 & 7297 &  940 & 6860 & 6870 \\
V097 & 0.334930  &  10720 &  4.54 & 0.20 & 7108 &  640 & 6805 & 6815 \\
V107 & 0.309031  &  10520 &  4.42 & 0.22 & 7259 &  880 & 6875 & 6880 \\
 \multicolumn{9}{l}{2nd overtone}\\
New var. & 0.249243 &  -- &  3.85 & 0.02 & 7755 &   90 & 7710 & 7710 \\
\hline  
&Median:&10480&&\\ 
&Mean:& 10480&&\\
&S.dev.:& 310&&\\
\hline
\multicolumn{9}{l}{$^a$$\Delta{R}$ is the peak to peak amplitude of the spectroscopic radius curve.}\\
\multicolumn{9}{l}{$^b$ $T_{\mathrm{min1}}$ is the absolute minimum value of the temperature variation.}\\
\multicolumn{9}{l}{$^c$ $T_{\mathrm{min2}}$ is the temperature value at the phase of light minimum.}
\end{tabular} 
\end{center}
\end{table*}

The other sources of the errors of the distance estimates (uncertainties of the $p$-factor, the synthetic colours, B.C., $T_{\mathrm{eff}}$-colour relation, reddening, zero points of the standard calibration of the photometry, etc.) are mostly systematic, and do not typically exceed $1-2$\,percent  \citep{error}.

Consequently, the random, uncorrelated errors account for about 5\,percent, while the systematic errors for $1-2$\,percent uncertainties of the distance estimates in M3. In principle, if the errors of the 26 individual distances  were perfectly uncorrelated, 
the accuracy of the mean distance were as good as 61\,pc.  Supposing that the systematic errors do not exceed 2\,percent, the accuracy of distance determination of M3 is thus estimated to be 210\,pc.

The average/median value of the derived distances of the 26 best, stable-light-cure RRL stars in M3 is 10480 pc, i.e. the distance modulus of the cluster is $\nu=15.10$\,mag.  For comparison, based on the pulsation properties of RRL stars, \cite{cassi} derived 10470 pc, and the \cite{h96} catalogue gives 10200 pc for the distance of M3. Using $K$-band  period-luminosity relations \cite{sollima} determined  $\nu=15.07$\,mag, modelling the pulsation light curves of RRL stars \cite{marconi} yielded $15.10\pm0.1$\,mag and fitting the colour-magnitude diagrams of the cluster adopting [Fe/H]$=-1.55$ and [$\alpha$/H]$=0.4$ zaro-age horizontal-branch models, isochrones and evolutionary tracks $\nu=15.05-15.04$\,mag distance modulus of M3 was obtained \citep{vb1,vb2}.

\subsubsection{The second step: the radii}

Fixing the distance of M3 to 10480 pc, the radius and the temperature variations of all the single-mode variables, which have reasonable RV  curves are determined according to the process described above, however,  only $R_0$  is determined in the fit this time.

 The fits of the $R_{\textrm{ph}}$ and $R_{\textrm{sp}}$ curves are shown in the right-hand panel of Fig.~\ref{std.bw} for 31 RRab, 6 first and 1 second-overtone variables assuming 10480 pc distance for each star. The differences between the matches of the $R_{\textrm{ph}}$ and $R_{\textrm{sp}}$ curves  according to the best fit and the fixed-distance solutions (left- and right-hand panels) are not perceivable.  With the exceptions of the shortest-period  overtone variables, V088, V086 and the second-overtone new variable, the match between the $R_{\textrm{ph}}$ and $R_{\textrm{sp}}$ curves is satisfactory for all the stars, even for those, which have only Hydra spectra or which RV curves are incomplete.

Phase differences between the $R_{\textrm{ph}}$ and $R_{\textrm{sp}}$ curves appear only for  V086 and the new variable. As simultaneous RV and photometric data of these stars are available,  phase inconsistency of the data cannot explain this. In the case of the new variable, we think that the problem arises from that the used grid, interpolated to 0.05 log$g$ and 5 K mesh, might not be accurate enough to follow the very small-amplitude changes of the $V-I_{\mathrm{C}}$ colour ($0.015-0.020$\,mag). However, this cannot be the explanation for V086, which star has a similar amplitude to other overtone variables.

The amplitude of the $R_{\textrm{ph}}$ curve of V088 is larger than the amplitude of its $R_{\textrm{sp}}$ curve. No simultaneous photometric and spectroscopic data are available for this star. However, as  no differences in the photometric amplitudes  are detected according to all the available CCD photometric observations obtained in the past 25 years, the possibility that real amplitude variation  would explain the differences of the $R_{\textrm{ph}}$ and $R_{\textrm{sp}}$ amplitudes is excluded. Both  the mean RV, the proper motion \citep{tuch}, and the mean magnitude values of V088  are consistent with  its cluster membership, so it is unlikely that its distance would differ significantly from the accepted 10480~pc value. Most probably, some defect of the Hydra spectra of this star results in an anomalously small amplitude of its $R_{\mathrm{sp}}$ curve.

The $4-8$th  columns of Table~\ref{bw} list the mean radius, the amplitude of the radius variation according to the spectroscopic data, the mean and the amplitude of $T_{\mathrm{eff}}$,  and the minimum  temperature values derived twofold. $T_{\mathrm{min1}}$ and $T_{\mathrm{min2}}$ correspond to the absolute minimum temperature value  occurring in the  $0.45-0.60$ pulsation-phase interval of RRab stars, and the temperature at light minimum, at phase $0.8-0.9$, respectively.

\section{Comparison with results of theoretical and  empirical studies}
\begin{figure}
\centering
\includegraphics[width=8.2cm]{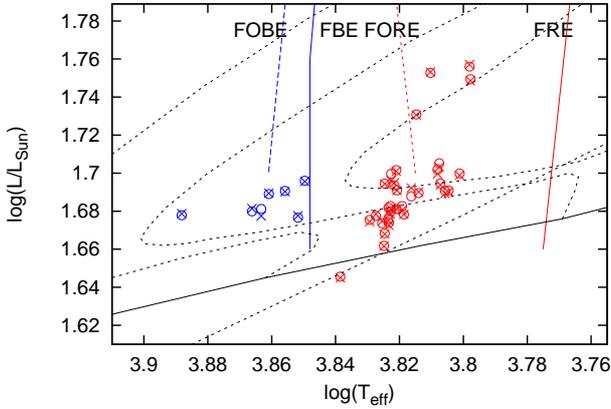}
\caption{The instability strip of M3, according to the data of the RR Lyrae stars that are the subject of the Baade-Wesselink analysis, is shown. Overtone and fundamental-mode variables are shown by blue and red symbols, respectively.  The luminosities, calculated using the two luminosity formulae, are shown by circles and crosses. The blue and red edges of the first-overtone and the fundamental-mode instability strips \citep[FOBE, FBE, FORE and FRE,][]{marconi} are also indicated by red and blue lines. The hottest star beyond the first-overtone blue edge is the new variable, which is probably a second-overtone RRL. The zero-age horizontal branch according to the $\alpha$ enhanced [Fe/H]$=-1.48$ models of \citet{dorman} and  the  [Fe/H]$=-1.5$ and [$\alpha$/Fe]$=0.4$ models of \citet{dotter} are drawn by continuous and dotted black lines, respectively.  The 0.62, 0.64 and 0.66 $M_\odot$ horizontal-branch evolutionary tracks of the Dorman models are also shown. } 
\label{hrd} 
\end{figure}

\subsection{The instability strip}

The instability strip occupied by the RRL stars of M3 is shown in Figure~\ref{hrd}, in comparison with model results of its boundaries according to the Z=0.001 models of \cite{marconi}. The luminosities of the stars are calculated twofold; the  $L/L_\odot={R/R_\odot}^2\times{T/T_\odot}^4$, and the $-2.5\mathrm{log}L/L_\odot=V_0+BC-5*\mathrm{log}d+5 - M_{\mathrm{Bol},\odot}$ formulae are applied. Only stars that are the subject of the BW analysis are shown. The $T_\mathrm{eff}$, $R$ and B.C. are determined in the course of the BW analysis, and the distance is taken to be 10480\,pc. 

The theoretical red boundary of both the fundamental and of the first overtone mode is at about 300\,K cooler temperature than defined by the sample of RRL stars shown. The hottest star observed, the new variable ($T_{\mathrm{eff}}=7755$\,K), lays at a 500\,K hotter temperature than the blue edge of the first-overtone instability region is, supporting that this star is a second-overtone RRL.
 
For comparison, the location of the zero-age horizontal branch according to appropriate composition stellar evolution models of \citet{dorman} and \citet{dotter} are also indicated in Fig.~\ref{hrd}.

\subsection{$A_{\mathrm{puls}} -A_V$ relation of RRL stars}\label{sect.a-a}
\begin{figure}
\centering
\includegraphics[width=8.2cm]{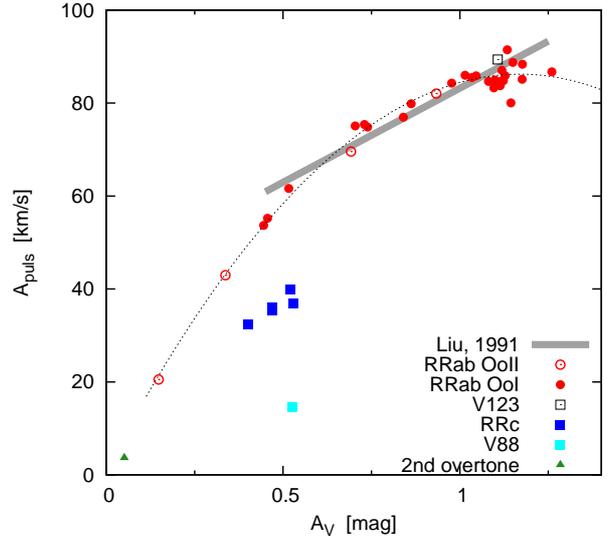}
\caption{Pulsation amplitude ($p*A_{\mathrm{RV}}$) versus the full amplitude of the $V$ light curve of RR Lyrae stars in M3 is compared with the $A_{\mathrm{puls}}-A_V$ relation determined by \citet{liu} (gray line) for Galactic field RRab stars.  \label{liu}}
\end{figure}

Based on  photometric and spectroscopic data,  \cite{liu} derived a linear relation between the pulsation amplitude ($A_{\mathrm{RV_{puls}}}= p A_{\mathrm{RV}}$; $p$:projection factor) and the $V$ amplitude of  stable-light-curve RRab stars. The $A_{\mathrm{RV}} - A_{V}$ relation of a Blazhko star in different phases of the modulation seems, however, to be  star by star  different and steeper than that of stable RRab stars \citep{distort}.

The  $A_{\mathrm{puls}} - A_{V}$ relation determined by \cite{liu}  in comparison with the M3 results is shown in Fig.~\ref{liu}. Liu's relation gives a very good linear approximation of the M3 data, however, covering  a larger amplitude (period) range, the relation is clearly nonlinear. The following formula
\begin{equation}
A_{\mathrm{puls}} = 145.46 A_{V} -62.22 {A^{2}}_{V}+1.21 
\end{equation}
\noindent 
fits the data of stable RRab stars in M3 with 2.2 km\,s$^{-1}$ rms scatter. It is important to note, that this relation is uniquely valid for OoI and OoII variables.

\subsection{Period-radius relation}

\begin{figure}
\centering
\includegraphics[width=8.2cm]{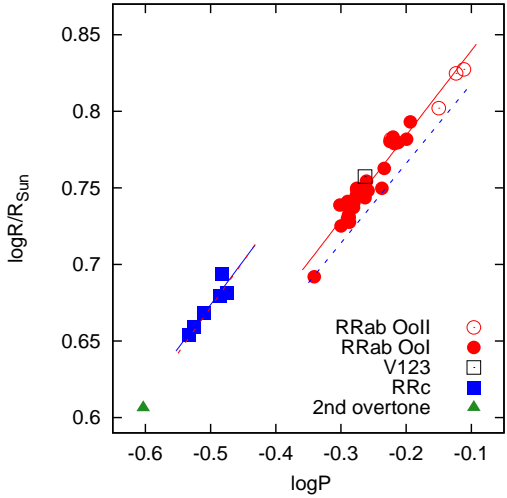}
\caption{Period -- radius relations of RR Lyrae stars in M3.  The linear fits  to the RRab and RRc data (continuous lines) and the $P-R$ relations derived by \citet{marconi} (dashed lines) are also shown. The two lines are practically identical for RRab stars and the offset between the empirical and theoretical relations of RRc stars is only about 0.015 in log$R$.}
\label{pr} 
\end{figure}

\begin{figure*}
\centering
\includegraphics[width=18.cm]{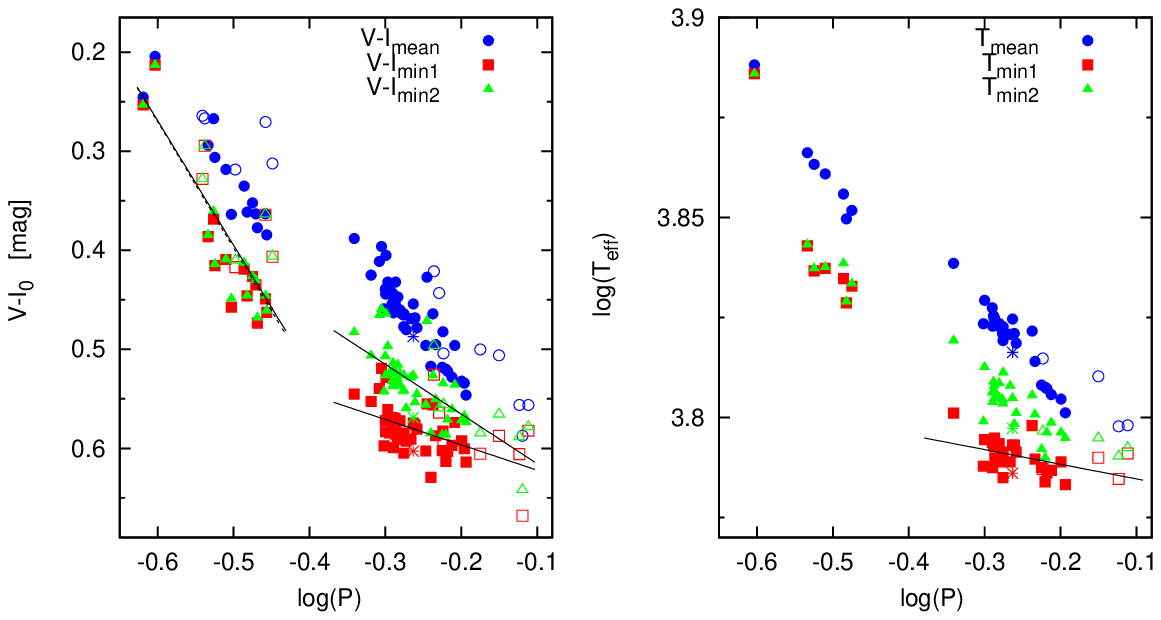}
\caption{Mean and  extrema of the  $(V-I)_0$ (maximum) and effective temperature (minimum) data of RR Lyrae stars.
The min1 and min2 denote the red/cold  extrema and the corresponding values   at light minimum, respectively.  The open symbols denote  OoII stars. All the stable-mode RR Lyrae stars with accurate $V$ and $I_{\mathrm C}$ photometric data are shown in the left-hand panel, and those which have both photometric and radial-velocity data are plotted in the right-hand panel. Note, that the values of the both the min1 and the min2 data are period dependent, and the period dependence of the min2 data of RRab stars is significantly larger than that of the min1 data. V123 is shown by star symbols.}
\label{tvi} 
\end{figure*}

The period dependence of the obtained mean radius values  of the variables are compared with model results in Fig.~\ref{pr}.
The period-radius (PR) relations defined by the fundamental and the first-overtone RRL stars and the theoretical relations for z=0.001 metallicity value \citep[eq. 7. and 8. in][]{marconi} are shown in this figure.  The fitted relations have the form of
\begin{align}
{\mathrm{log}}(R/R_\odot) &= 0.872(\pm0.006) +  0.555(\pm0.025) {\mathrm{log}} P_{\mathrm{FU}},\label{prfu}\\
{\mathrm{log}}(R/R_\odot) &= 0.94(\pm0.06) +  0.57(\pm0.12) {\mathrm{log}} P_{\mathrm{FO}}\label{prfo}
\end{align}
\noindent for the fundamental (FU) and the first overtone (FO) variables, respectively. 

The agreement between the theoretical and the observed PR relations is excellent; both the slopes and the zero points (if corrected for the metallicity dependence) of the relations are the same within the $1\,\sigma$ limit for both the RRab and the RRc stars. Although the theoretical log$R$ values seem to be larger by 0.015 than the radii determined from the BW analysis, this difference remains within the uncertainty limit, moreover, such a difference can arise e.g., from a minor, $\sim0.10$ offset of the log$g$ values accepted for the overtone variables.

\subsection{Minimum colour and temperature values}

Fig.~\ref{tvi} plots the period dependence of the mean and the min1 and min2  values of the dereddened  $(V-I)_{0}$ and $T_{\mathrm{eff}}$ (see Table~\ref{bw}) of non-Blazhko and non-double-mode RRL stars in M3.
 The min1 corresponds to the red (cold) extreme of the colour (temperature) again, while min2 denotes their values  at light minimum.

While the mean colours and the temperature of RRL stars  depend on the period of the pulsation strongly, the  colours and the temperature at light minimum have been supposed to be  nearly the same at any period value especially for fundamental-mode variables. \cite{guld} derived a $(V-I)_{\mathrm{0,min}} - $[Fe/H] relation using the average value of the $(V-I)_{0}$ over the $0.5-0.8$ pulsation-phase range for Galactic field RRL stars. This relation  yields  $(V-I)_{\mathrm{0,min}}$=0.58 mag value for the metallicity of M3. The  mean of the $(V-I)_{\mathrm{0,min1}}$  values of the RRab stars in M3 is 0.583, which is  in excellent agreement with this result.

Studying the period-colour relations of RRL stars using the OGLE III data \citep{so09,so10} of the Large and Small Magellanic Clouds (MCs), \cite{kanbur} found that the slope of the $(V-I)_0$ versus log$P$ relation for RRab and RRc stars are 0.093/0.055 and 0.604/0.472 in the LMC/SMC, respectively. $(V-I)_0$ corresponds the min2 data here, measured at the phase of the light minimum.
The linear fits to the RRab data shown in the left-hand panel of Fig.~\ref{tvi} have, however, significantly larger,  0.502 and 1.26 slopes for the min2 data for the fundamental and the overtone variables, respectively. These are  $2-10$ times larger than the corresponding values determined for the data of the MCs. Although the period dependence of the min1 data of RRab stars is significantly smaller than for min2, its slope is still not negligible for the $(V-I)_{\mathrm{0,min1}}$ data.  The form of the linear fits to the min1 data of the $(V-I)_0$ and the $T_{\mathrm{eff}}$ data of RRab stars shown in Fig.~\ref{tvi}  are the followings:
\begin{align}
(V-I)_{\mathrm{0,min1}} &= 0.65(\pm0.02)   +  0.26(\pm0.06)  {\mathrm{log}} P_{\mathrm{FU}},\label{vifu}\\
 T_{\mathrm{eff,min1}} &= 3.781(\pm0.003) +  0.037(\pm0.013) {\mathrm{log}} P_{\mathrm{FU}}.\label{tfu}
\end{align}
Concerning the overtone variables, the period dependences of the $(V-I)_0$ values using min1 and min2 data are practically the same; their steepness is 1.26. 

Clearly, the relative constancy of the minimum temperature and the $(V-I)_0$ colour of RRL stars is valid only for the fundamental mode and for the absolute minimum values (min1),  while the mean and the minimum-light colors are strongly period  dependent  if a homogeneous sample of variables, like in M3, is investigated.

The discrepancy between the detected period dependencies of the minimum colours in M3 and in the MCs arises, most probably, from that  the scarceness of the $V$ light curves and the larger uncertainties of the photometric data hinder to determine the $(V-I)_{\mathrm{0,min}}$ values in the MCs accurately enough. 

\section{V123}

\begin{figure}
\centering
\includegraphics[width=8.6cm]{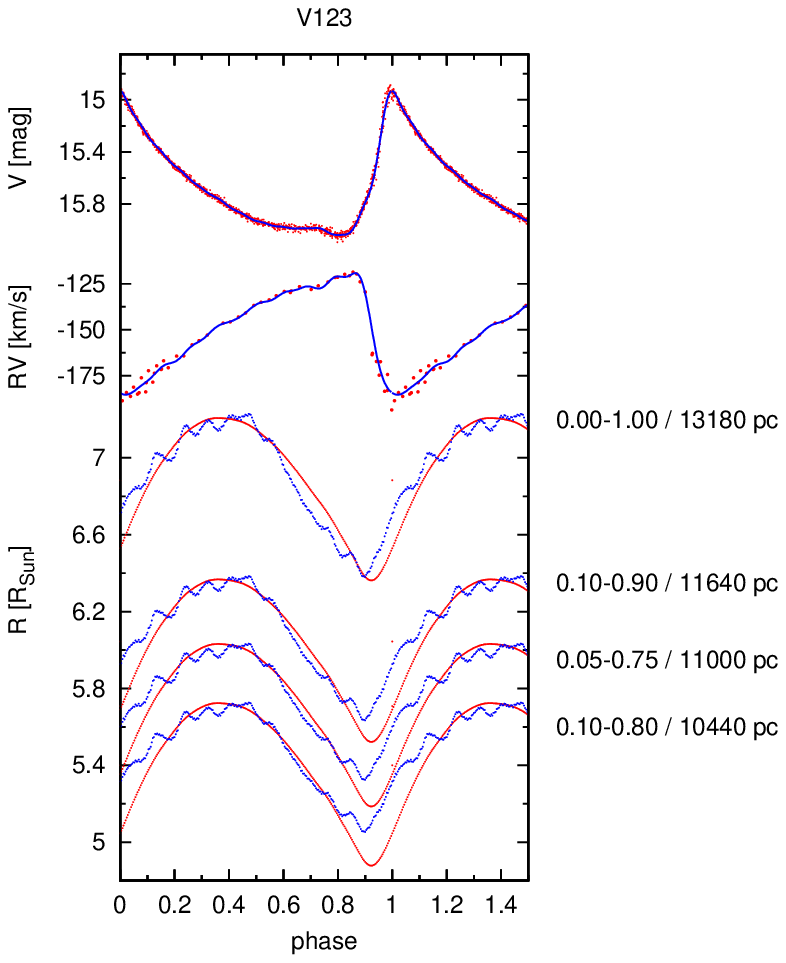}
\caption{The observed and the fitted $V$ and radial-velocity curves and the $R_{\mathrm {sp}}$ and $R_{\mathrm {phot}}$ variations according to different Baade-Wesselink solutions are shown for V123. Different  distances are derived if different phase intervals of the $R_{\mathrm {sp}}$ and $R_{\mathrm {phot}}$ curves are concerned in the fitting process. The phase interval of the fit, and the derived distances  are given in right side of the panel for some solutions. }
\label{v123} 
\end{figure}

The peculiar light-curve shape of V123 was discussed in detail in \cite{v123}. Based on its brightness, colours and amplitudes, V123 is similar to RRab stars of the same period in M3, but its light-curve has a smaller bump and a longer and less steep rising branch than other, stable-light-curve RRab stars have. Although the light curve resembles to the light curves of some Blazhko stars, no clear evidence of the Blazhko effect has been detected; only very small-amplitude stochastic variation around maximum brightness is evident. 

In spite of that the position of V123 was a bit offset in the Hectochelle@MMT observations and the spectra have low S/N,  reasonable RV data of these spectra could also be obtained. The scatter of this RV curve is larger than for the other RRL stars, however, this is the consequence of the low quality of the spectra, and does not indicate real RV variations.

In order to find an answer to the peculiarities of V123, a complete BW analysis of this star is also performed. The observed and fitted $V$ and RV curves and the  matching of the $R_{\mathrm {sp}}$ and $R_{\mathrm {phot}}$ radius curves using different phase intervals for the fit are shown in Fig.~\ref{v123}. The phase intervals involved in the fitting process and the derived distances are given in the right-hand side of the panel for the BW solutions shown. If the phase interval between 0.8 and 1.1 is omitted, the distance derived for V123  is 10440 pc, in perfect agreement with the mean distance value obtained for M3. If smaller or no phase interval is skipped, then the distance become larger, and a systematic difference between the radius curves emerges in the  $0.6-0.8$ phase interval. In these solutions, the $R_{\mathrm {sp}}$ is larger than $R_{\mathrm {phot}}$ all along the $0.6-0.8$ phases, which is not the case for any of the other RRab stars analysed as it can bee seen in Fig.~\ref{std.bw}. 

Therefore, the BW analysis  of V123 seems to strengthen the conclusion that the star is indeed a cluster member. The $R_{\mathrm {sp}}$ and $R_{\mathrm {phot}}$ curves of V123 using the adopted 10480 pc distance of the cluster is shown in the right-hand panel of Fig.~\ref{std.bw}. The matching of the two radius curves is similarly good as for other RRab stars between phases 0.15 and 0.80, however, the discrepancy between the two curves is the largest in the $0.80-1.15$ phase interval in the whole sample of the stars analysed.   Typically, $R_{\mathrm {phot}}$ is $0.1-0.2$~$R_\odot$ larger than $R_{\mathrm {sp}}$ between phases 0.9 and 1.1 for the shorter period, large-amplitude RRab stars, i.e. in those stars which develop large atmospheric shocks. In the case of V123, the discrepancy of the radius curves affects a longer phase interval  and its amplitude is also larger ($0.2-0.3$~$R_\odot$) than in the other stars.

The large discrepancy of the radius curves around the rising branch of the light curve indicates strong anomalies of the atmosphere in these phases in line with that the biggest discrepancy in the shape of the light curve appears also in this phase interval. Most probably, an anomalous, strong shock, appearing at a peculiar phase, is behind the anomalies of V123.

The positions of V123 are shown by special symbols in the different figures of the present paper.
It was already shown in \cite{v123} that the $R_{k1}$ ($k>2$) Fourier amplitude ratios and the higher-order phase differences $\varphi_{k1}$ $(k>5)$ of V123 are anomalously small, as it is also documented in Fig.~\ref{four}. This is in contrast with the behaviour of the Fourier parameters of the RV curve. Only the $A_5$, $A_6$, $A_7$ amplitudes and the $R_{51}$, $R_{61}$ and $R_{71}$ amplitude ratios are smaller than normal, but no significant anomaly of the phase differences of the RV curve is evident. This is in line with that the shape of the  RV curve does not look to be so peculiar as the shape of the light curve is (see Fig.~\ref{v123}). 

The $A_{\mathrm{puls}} -A_V$ relation, the radius and the minimum colours and temperatures of V123 are normal, as documented in Fig.~\ref{liu},~\ref{pr} and \ref{tvi}. It seems that, with the exception of its light-curve shape, all the properties of V123 are similar to the properties of a normal RRab star.

\section{Summary}\label{sect.sum}

Radial velocity and photometric data of a large set of RRL variables in M3 are published.
This is the first extended RV survey of RRL stars in a GC. Moreover, this is the largest set of RV curves of RRL stars ever obtained.

The main results derived from these data are as follows.
\begin{itemize}

\item
$BVI_{\mathrm C}$ magnitude of 111 and instrumental $B$-band flux time series of 64 RRab stars are presented. The light curves of 47\,percent of the variables show the Blazhko effect.

\item
The mean value of the $\gamma$-velocities of  79 RRL stars is $-146.8$\,km\,s$^{-1}$ with 4.52\,km\,s$^{-1}$ standard deviation. The velocity dispersion of RRL stars at different projected radial distances ($r<12$ arcmin) are slightly but systematically smaller  than the velocity dispersion of giants, however, the differences are not statistically significant.

\item 
Difference of about $1-2$\,km\,s$^{-1}$  between the mean RVs of Blazhko stars in different phases of the modulation are detected. This is  probably a consequence of the dynamics of the atmosphere and/or the reduction process, and does not reflect real changes of the $\gamma$-velocities.

\item 
The  ${\Phi_{21}}^{\mathrm RV}$ phase differences of the RV curves of RRab stars seem to be uniformly constant. Comparing the  ${\Phi_{21}}^{\mathrm RV}$ values of  M3 and Galactic field stars, no period or metallicity dependence, larger than the uncertainty would indicate, has been found.

\item
The mean value of the BW distances of 26 single-mode RRL stars is $10480\pm210$ pc, which is in good agreement with literature data.

\item 
The $A_{\mathrm{puls}}-A_V$ relation of RRab stars is quadratic, and OoI and OoII stars obey the same relation. 

\item
The period - radius relation defined by the results of the BW analysis of 6 RRc and 31 RRab stars are in very good agreement with theoretical model results \citep{marconi}.

\item
The $(V-I)_0$ colour and the temperature at the phase of minimum light  is found to show a clear period dependence; the differences between the shortest and longest period RRab stars are about  0.1 mag and 500 K. The  period dependence of the absolute minimum value  of the temperature (the absolute maximum values of the colours), which occur at pulsation phase between 0.45 and 0.60, is less significant. The mean values of the extrema of the $(V-I)_0$ and temperature data of non-Blazhko RRab stars in M3 are $0.583\pm0.025$ mag and  $6170\pm60$ K, respectively.

\end{itemize}

A detailed analysis of the photometry and RV data of Blazhko stars is going to be published in forthcoming papers.

\section*{Acknowledgments}
The constructive comments of the referee, Antonio Sollima, helped to inprove the paper.
GH acknowledges support by the Ministry for the Economy, Development, and Tourism's Programa Iniciativa Cient\'ifica Milenio through grant IC210009, awarded to the Millennium Institute of Astrophysics; by Proyecto Basal PFB-06/2007; by Fondecyt grant 1141141; and by CONICYT-PCHA/Doctorado Nacional grant 2014-63140099. \'AS and KV were supported by the J\'anos Bolyai Research Scholarship of the HAS and PS by PECS-98073. KK gratefully acknowledges the support of a Marie Curie IOF under FP7 (PIOF-255267 SAS-RRL). CAP gratefully acknowledges support from the Daniel Kirkwood endowment at Indiana University. NJ acknowledges the NKFIH K-115709 grant, MA the grant of the MTA CSFK Lend\"ulet Disk Research Group, AP the OTKA K-113117 and LP2012-31 grants. KS has been supported by the Hungarian OTKA grant K-104607.


\begin{thebibliography}{99}
\bibitem[Alard(2000)]{isis} Alard C. 2000, A\&A Suppl., 144, 363
\bibitem[Benk\H{o} et al.(2006)]{be06} Benk\H{o} J. M., Bakos G. \'A.  Nuspl, J. 2006, MNRAS, 372, 1657 
\bibitem[\protect\citeauthoryear{Bhardwaj et al.}{2014}]{kanbur} Bhardwaj A., Kanbur S. M., Singh H. P., Ngeow C.-C.  2014, MNRAS, 445, 2655
\bibitem[Cacciari et al.(1992)]{ccf} Cacciari C., Clementini G., Fernley J. A. 1992, ApJ, 396, 219
\bibitem[Cacciari et al.(2005)]{ca05} Cacciari C., Corwin T.M., Carney B.W. 2005, AJ, 129, 267
\bibitem[Carney et al.(1992)]{c92} Carney B. W., Storm J., Jones R. V. 1992, ApJ, 386, 663
\bibitem[\protect\citeauthoryear{Carney}{1996}]{carney}  Carney B. W. 1996, PASP, 108, 900
\bibitem[\protect\citeauthoryear{Cassisi et al.}{2001}]{cassi} Cassisi S., De Santis R., Piersimoni A. M. 2001, MNRAS, 326, 342
\bibitem[Castellani et al.(2005)]{ccc} Castellani M., Castellani V., Cassisi S. 2005 A\&A, 437, 1017
\bibitem[Castelli \& Kurucz(2003)]{kurucz} Castelli F.  Kurucz R. L. 2003, in Modelling of Stellar Atmospheres, ed. N. Piskunov, W.-W. Weiss and D.-F. Gray, IAU Symp., 210, 20
\bibitem[Catelan(2004)]{ca4} Catelan, M. 2004 ApJ, 600, 409
\bibitem[Chadid \& Preston(2013)]{cp} Chadid, M., Preston G. W. 2013, MNRAS, 434, 552
\bibitem[Cohen(1992)]{cohen} Cohen J. G. 1992, ApJ, 400, 528
\bibitem[Cohen \& Mel\'endez(2005)]{cm} Cohen J. G., Mel\'endez J. 2005, AJ, 129, 303
\bibitem[Clementini et al.(2004)]{cc04} Clementini G., Corwin T. M., Carney B. W.,   Sumerel A.N. 2004, AJ, 127, 938 
\bibitem[Clementini et al.(1990)]{cc90} Clementini G., Cacciari C., Lindgren H. 1990, A\&A Suppl., 85, 865
\bibitem[Corwin \& Carney(2001)]{co01} Corwin B. W.,  Carney B. W. 2001, AJ, 122, 3183 
\bibitem[Dorman(1992)]{dorman} Dorman B., 1992, ApJ Suppl., 81, 221
\bibitem[Dotter et al.(2007)]{dotter} Dotter A., Chaboyer B., Jevremovi\'c D., Baron E., Ferguson J. W., Sarajedini A., Anderson J. 2007, AJ, 134, 376	
\bibitem[Feuchtinger(1999)]{feu} Feuchtinger M. U. 1999, A\&A, 351, 103
\bibitem[Fernley(1994)]{f94} Fernley J.,  1994, A\&A, 284, 16
\bibitem[Gautschy(1987)]{gautschy} Gautschy A. 1987, Vistas in Astronomy, 30, 197
\bibitem[Gillet \& Fokin(2014)]{gf} Gillet, D., \&   Fokin, A. B. 2014, A\&A, 565, 73
\bibitem[\protect\citeauthoryear{Givens \& Pilachowski}{2016}]{gp} Givens R. A., Pilachowski C. A.  2016, PASP, 128, 4203
\bibitem[\protect\citeauthoryear{Guldenschuh et al.}{2005}]{guld} Guldenschuh K. A. et al. 2005, PASP, 117, 721
\bibitem[\protect\citeauthoryear{Gunn \& Griffin}{1979}]{gg} Gunn J. E., Griffin R. F. 1979, AJ, 84, 752
\bibitem[\protect\citeauthoryear{Harris}{1996}]{h96} Harris W. E. 1996 (2010 edition), AJ, 112, 1487
\bibitem[Hartman et al.(2005)]{H05} Hartman J. D., Kaluzny J., Szentgy\"orgyi A.,  Stanek K. Z. 2005, AJ, 129, 1596 
\bibitem[\protect\citeauthoryear{Hernandez \&  Jim\`enez}{2012}]{hernandez} Hernandez X., Jim\`enez M. A. 2012, ApJ, 750, 9
\bibitem[Johnson et al.(2005)]{j05} Johnson C. I., Kraft R. P., Pilachowski C. A., Sneden C., Ivans I. I., Benman G.
2005, PASP, 117, 1308
\bibitem[Jones et al.(1992)]{jcsl} Jones R. V., Carney B. W., Storm J., Latham D. W. 1992, ApJ, 386, 646
\bibitem[Jurcsik(1998)]{j98} Jurcsik J., 1998, A\&A, 333, 571
\bibitem[Jurcsik \& Kov\'acs(1996)]{jk96} Jurcsik J.,  Kov\'acs, G. 1996, A\&A, 312, 111
\bibitem[Jurcsik et al.(2002)]{distort} Jurcsik J., Benk\H o, J., Szeidl, B. 2002, A\&A, 390, 133
\bibitem[Jurcsik et al.(2012)]{oc} Jurcsik J., Hajdu G., Szeidl B. et al. 2012, MNRAS, 419, 2173 
\bibitem[Jurcsik et al.(2013)]{v123} Jurcsik J. et al 2013, ApJ Letters, 776, L1
\bibitem[Jurcsik et al.(2015)]{overtone} Jurcsik J. et al. 2015, ApJ Suppl. 219, 25
\bibitem[Jurcsik \& Smitola(2016)]{rrl} Jurcsik J., Smitola P. 2016, Communications from the Konkoly Observatory, Vol. 105, p. 167
\bibitem[Kaluzny et al.(1998)]{K98} Kaluzny J., Hilditch R. W., Clement C.,  Rucinski S. M. 1998, MNRAS, 296, 347 
\bibitem[Kamann et al.(2014)]{Kam2000} Kamann, S. et al., 2014, A\&A, 566, A58
\bibitem[\protect\citeauthoryear{Kimmig et al.}{2015}]{kimmig} Kimmig B., Seth A., Ivans I. I., Strader J., Caldwell N., Anderton T., Gregersen D. 2015,  AJ, 149, 53
\bibitem[King(1966)]{king} King I.R. 1966, AJ 71,64
\bibitem[Koll\'ath(1990)]{mufran} Koll\'ath Z. 1990, Occ. Techn. Notes Konkoly Obs., No. 1, http://www.konkoly.hu/staff/kollath/mufran.html
\bibitem[\protect\citeauthoryear{Kov\'acs}{2003}]{k03} Kov\'acs G. 2003, MNRAS, 342, L58
\bibitem[\protect\citeauthoryear{Kov\'acs \& Walker}{2001}]{kw} Kov\'acs G., Walker A. R. 2001, A\&A, 371, 579
\bibitem[\protect\citeauthoryear{Liu}{1991}]{liu} Liu T. 1991, PASP, 103, 205
\bibitem[Liu \& Janes(1990a)]{lj90} Liu T., Janes K. A. 1990, ApJ, 354, 273
\bibitem[\protect\citeauthoryear{Liu \& Janes}{1990b}]{liu90} Liu T., Janes K. A., 1990, ApJ, 360, 561
\bibitem[\protect\citeauthoryear{Marconi et al.}{2015}]{marconi} Marconi M et al. 2015, ApJ, 808, 50
\bibitem[\protect\citeauthoryear{Marconi et al.}{2003}]{mar03} Marconi M., Caputo F., Di Criscienzo M., Castellani M. 2003, ApJ, 596, 299
\bibitem[Marconi \& Degl'Innocenti(2007)]{marc} Marconi M., Degl'Innocenti S. 2007, A\&A, 474, 557
\bibitem[\protect\citeauthoryear{Marengo et al.}{2004}]{error} Marengo M., Karovska M., Sasselov D. 2004, ApJ, 603, 285
\bibitem[Moskalik et al.(2000)]{m00} Moskalik P., Krzyt T., Gorynya N. A., Samus N. N. 2000, in The Impact of Large-Scale Surveys on Pulsating Star Research, ed. L. Szabados and D. Kurtz, ASPC, 203, 233
\bibitem[Moskalik (2014)]{mo14} Moskalik P. 2014, in Precision Asteroseismology, ed., J.A. Guzik, W.J. Chaplin, G. Handler \& A. Pigulski, Proceedings IAU Symposium No. 301, 249
\bibitem[\protect\citeauthoryear{Nardetto et al.}{2004}]{n04} Nardetto N., Fokin A., Mourard D., Mathias Ph., Kervella P., Bersier D. 2004, A\&A, 428, 131 
\bibitem[\protect\citeauthoryear{Navin et al.}{2016}]{n16}  Navin C. A.,  Martell S. L.,  Zucker D. B. 2016, ApJ, 829, 123
\bibitem[Pilachowski et al.(2000)]{Pil2000} Pilachowski C. A., Sneden C., Kraft R.P., Harmer D., Willmarth D. 2000, AJ, 119, 2895
\bibitem[Pont et al.(2001)]{pont} Pont F., Kienzle F., Gieren W., Fouqu\'e P. 2001, A\&A, 376, 892
\bibitem[Preston(2011)]{preston}  Preston G. W. 2011, AJ, 141, 6
\bibitem[Samus et al.(2009)]{samus} Samus N. N., Kazarovets E. V., Pastukhova E. N., Tsvetkova T. M., Durlevich O. V. 2009, PASP, 121, 1378
\bibitem[Sandstrom et al.(2001)]{san} Sandstrom K., Pilachowski C. A.,  Saha A. 2001, AJ, 122, 3212
\bibitem[Simon \& Teays(1982)]{st} Simon N. R., Teays T. J. 1982, ApJ, 261, 586
\bibitem[Skillen et al.(1989)]{sk89} Skillen I., Jameson R. F., Fernley J. A., Lynas-Gray A. E., Longmore A. J. 1989, MNRAS, 241, 281
\bibitem[Skillen et al.(1993)]{sk93} Skillen I., Fernley J. A., Stobie R. S., Jameson R. F. 1993, MNRAS, 265, 301
\bibitem[\protect\citeauthoryear{Smart \& Nicastro}{2014}]{smart} Smart R. L., Nicastro L. 2014, A\&A, 570, 87
\bibitem[Smolinski et al.(2011)]{Smol2011} Smolinski J.P., Martell S.L., Beers T.C., Lee Y. S. 2011, AJ, 142, 126
\bibitem[Sneden et al.(2004)]{sneden} Sneden C., Kraft R. P., Guhathakurta P., Peterson R.C., Fulbright J. P. 2004, AJ, 127, 2162
\bibitem[Soderberg et al.(1999)]{sod} Soderberg A. M., Pilachowski C. A., Barden S. C., Willmarth D., Sneden C. 1999, PASP, 111, 1233
\bibitem[S\'odor (2012)]{nl} S\'odor \'A. 2012 Occ. Techn. Notes Konkoly Obs., No. 1, http://www.konkoly.hu/staff/sodor/lcfit.html
\bibitem[Sollima et al.(2006)]{sollima} Sollim, A., Cacciari C., Valenti E. 2006, MNRAS, 372, 1675
\bibitem[\protect\citeauthoryear{Soszy\'nski et al.}{2009}]{so09} Soszy\'nski I. et al. 2009, AcA, 59, 1 
\bibitem[\protect\citeauthoryear{Soszy\'nski et al.}{2010}]{so10} Soszy\'nski I. et al. 2010, AcA, 60, 165 
\bibitem[Stetson(2000)]{st00} Stetson, P. B. 2000, PASP, 112, 925
\bibitem[\protect\citeauthoryear{Storm et al.}{1994a}]{storm1} Storm J., Carney B. W., Latham D. W. 1994, A\&A, 290, 443 
\bibitem[Storm et al.(1994b)]{st94} Storm J., Carney B. W., Nordstr\"om B., Andersen J., Latham D. 1994, in Hot Stars in the Galactic Halo, ed. Saul J. Adelman and Arthur R. Upgren. Cambridge: Cambridge University Press,  p.298
\bibitem[\protect\citeauthoryear{Storm et al.}{1994c}]{storm2} Storm J., Nordstrom B., Carney B. W., Anderson J.  1994, A\&A, 291, 121
\bibitem[Szeidl(1988)]{szeidl} Szeidl B., 1988, in Multimode Stellar Pulsation, ed. G. Kov\'acs, L. Szabados \& B. Szeidl (Budapest), p51
\bibitem[Szentgy\"orgyi  et al.(2011)]{mmt} Szentgy\"orgyi A., F\H{u}r\'esz G., Cheimets P. et al. 2011, PASP, 123, 1188
\bibitem[Tonry \& Davis(1979)]{ton} Tonry J., Davis M. 1979, AJ, 84, 1511
\bibitem[Torres(2010)]{torres} Torres G. 2010, AJ, 140, 1158
\bibitem[Tucholke et al.(1994)]{tuch} Tucholke H.-J., Scholz R.-D., Brosche P. 1994, A\&AS, 104, 161
\bibitem[VandenBerg et al.(2016)]{vb2} VandenBerg D. A., Denissenkov P. A., Catelan M. 2016, ApJ, 827, 2
\bibitem[VandenBerg et al.(2015)]{vb1} VandenBerg D. A., Stetson P. B., Brown T. M. 2015, ApJ, 805, 103
\end{thebibliography}
\end{document}